\documentclass[12pt]{jnmp}

\usepackage{amsmath,amssymb,amsthm,amsfonts}

\usepackage{array}

\usepackage{latexsym,epsfig}

\numberwithin{equation}{section}

\DeclareMathOperator{\rank}{rank}
\DeclareMathOperator{\grad}{grad}
\DeclareMathOperator{\dive}{div}
\newcommand{\tr}{Tr}

\newtheorem{theo}{Proposition}

\newcommand{\AB}{\allowbreak}

\newcommand{\ali}[2]{\mathop{\mathfrak{#1}(#2)}\nolimits}

\newcommand{\ad}{\mathop{\mathrm{ad}}\nolimits}

\newcommand{\ADA}[1]{\ifmmode \ad(#1) \else $\ad(#1)$\fi}

\newcommand{\LI}[2]{\ifmmode#2_1,\AB\,\ldots,\,\AB #2_{#1}%

\else$ #2_1,\AB\,\ldots,\,\AB#2_{#1}$\fi}

\newcommand{\sltwo}{\ifmmode \ali{sl}{2} \else $\ali{sl}{2}$\fi}

\newcommand{\bMA}[1]{\[\begin{array}{#1}}

\newcommand{\eMA}{\end{array}\]}

\def\be{\begin{equation}}

\def\ee{\end{equation}}

\def\p{{\partial}}

\def\tr{{\mathrm{Tr}}}

\mathchardef\za="710B  %\alpha

\mathchardef\zb="710C  %\beta

\mathchardef\zg="710D  %\gamma

\mathchardef\zd="710E  %\delta

\mathchardef\ze="710F  %\epsilon

\mathchardef\zz="7110  %\zeta

\mathchardef\zh="7111  %\eta

\mathchardef\zy="7112  %\theta

\mathchardef\zi="7113  %\iota

\mathchardef\zk="7114  %\kappa

\mathchardef\zl="7115  %\lambda

\mathchardef\zm="7116  %\mu

\mathchardef\zn="7117  %\nu

\mathchardef\zx="7118  %\xi

\mathchardef\zp="7119  %\pi

\mathchardef\zr="711A  %\rho

\mathchardef\zs="711B  %\sigma

\mathchardef\zt="711C  %\tau

\mathchardef\zu="711D  %\upsilon

\mathchardef\zf="711E  %\phi

\mathchardef\zq="711F  %\chi

\mathchardef\zc="7120  %\psi

\mathchardef\zw="7121  %\omega

\mathchardef\zG="7000  %\Gamma

\mathchardef\zD="7001  %\Delta

\mathchardef\zY="7002  %\Theta

\mathchardef\zL="7003  %\Lambda

\mathchardef\zX="7004  %\Xi

\mathchardef\zP="7005  %\Pi

\mathchardef\zS="7006  %\Sigma

\mathchardef\zU="7007  %\Upsilon

\mathchardef\zF="7008  %\Phi

\mathchardef\zC="7009  %\Psi

\mathchardef\zW="700A  %\Omega

\mathchardef\ze="7122  %\varepsilon

\mathchardef\zvy="7123  %\vartheta

\mathchardef\zvr="7125 %\varrho

\mathchardef\zvs="7126 %\varsigma

\mathchardef\zvf="7127  %\varphi

\title{Riemann Invariants and Rank-k Solutions of Hyperbolic Systems}

\bigskip

\author{
A.M. Grundland
\\
Centre de Recherches Math{\'e}matiques, Universit{\'e} de Montr{\'e}al, \\
C. P. 6128, Succ.\ Centre-ville, Montr{\'e}al, (QC) H3C 3J7, Canada \\
Universit{\'e} du Qu{\'e}bec, Trois-Rivi{\`e}res CP500 (QC) G9A 5H7, Canada \\
email address : grundlan@crm.umontreal.ca
\\\\\\\
B. Huard\\
D{\'e}partement de math{\'e}matiques et de statistiques,\\
C.P. 6128, Succc.\ Centre-ville, Montr{\'e}al, (QC) H3C 3J7, Canada \\
email address : benoit.huard@dms.umontreal.ca
}
\begin{document}

\Name{Rank-k Solutions Described by Hyperbolic Systems}

\Author{A.M. Grundland $^\dag$ and B. Huard $^\ddag$}

\Address{$^\dag$ Centre de Recherches Math{\'e}matiques, Universit{\'e} de Montr{\'e}al, \\
C. P. 6128, Succ.\ Centre-ville, Montr{\'e}al, (QC) H3C 3J7, Canada \\
Universit{\'e} du Qu{\'e}bec, Trois-Rivi{\`e}res CP500 (QC) G9A 5H7, Canada \\
email address : grundlan@crm.umontreal.ca\\[10pt]
$^\ddag$ D{\'e}partement de math{\'e}matiques et de statistiques,\\
C.P. 6128, Succc.\ Centre-ville, Montr{\'e}al, (QC) H3C 3J7, Canada \\
email address : benoit.huard@dms.umontreal.ca}

%\maketitle

%\newpage
\begin{abstract}

In this paper we employ a "direct method" in order to obtain
rank-k solutions of any hyperbolic system of first order
quasilinear differential equations in many dimensions.  We discuss
in detail the necessary and sufficient conditions for existence of these type
of solutions written in terms of Riemann invariants.  The most
important characteristic  of this approach is the introduction of
specific first order side conditions consistent with the original
system of PDEs, leading to a generalization of the Riemann
invariant method of solving multi-dimensional systems of PDEs.  We
have demonstrated the usefulness of our approach through several
examples of hydrodynamic type systems; new classes of solutions have been obtained in a closed form.

\begin{center}
{\bf R{\'e}sum{\'e}}
\end{center}
Dans cet article, nous employons une "m{\'e}thode directe" pour obtenir des solutions de rang k pour tout syst{\`e}me
hyperbolique d'{\'e}quations diff{\'e}rentielles quasilin{\'e}aires du premier ordre en plusieurs dimensions.  Nous
discuterons en d{\'e}tail les conditions n{\'e}cessaires et suffisantes pour l'existence de ces types de solutions {\'e}crites en
termes d'invariants de Riemann.  La caract{\'e}ristique la plus importante de cette approche est l'introduction de
contraintes diff{\'e}rentielles de premier ordre suppl{\'e}mentaires et compatibles avec le syst{\`e}me d'EDPs original, ce
qui conduit {\`a} une g{\'e}n{\'e}ra-lisation de la m{\'e}thode des invariants de Riemann pour la r{\'e}solution de
syst{\`e}mes d'EDPs en plusieurs dimensions.  Nous avons d{\'e}montr{\'e} l'utilit{\'e} de cette approche par plusieurs
exemples de syst{\`e}mes de type hydrodynamique et des nouvelles classes de solutions sont obtenues.\\

PACS subject classification (1994) : Primary 03.40.Kf; Secondary 02.20.Sv

\end{abstract}
\medskip

%Keywords and phrases: Sigma models, complex projective space, Lie

%groups, Lie algebras, differential geometry of surfaces.

\newpage

\section{Introduction}

\label{sec:Intro}
\text{}

This work has been motivated by a search for new ways of constructing multiple Riemann waves for nonlinear hyperbolic systems.  Riemann waves and their superpositions were first studied two
centuries ago in connection with differential equations describing
a compressible isothermal gas flow, by D. Poisson \cite{Poi1} and
later by B. Riemann \cite{Rie1}. Since then many different
approaches to this topic have been developed by various authors
with the purpose of constructing solutions to more
general hydrodynamic-type systems of PDEs.  For a classical
presentation we refer reader to a treatise by R. Courant and D. Hilbert
\cite{Cou1} and for a modern approach to the subject, see e.g.
\cite{Joh1,Per2,Roz1} and references therein.  A review
of most recent
developments in this area can be found in \cite{Daf1,Dub1,Maj1}.\\

The task of constructing multiple Riemann waves has been approached so far through the method of characteristics.  It relies on treating Riemann invariants as new independent variables (which remain constant along appropriate characteristic curves of the basic system).  This leads to the reduction of the dimensionality of the initial system which has to be subjected however to the additional differential constraints, limiting the scope of resulting solutions.

We propose here a new (though a very natural) way of looking at solutions expressible in terms of Riemann invariants, namely from the point of view of their group invariance properties.  We show that this approach (initiated in \cite{Doy1,Gru4}) leads to the larger classes of solutions, extending beyond Riemann multiple waves.

We are looking for the rank-k solutions of first order quasilinear hyperbolic system of PDEs in $p$
independent
variables
$x^i$ and $q$ unknown functions $u^{\alpha}$ of the form
\begin{equation}
\label{EQ1-1}
\Delta^{\mu i}_{\,\,\alpha}(u)\,u^{\alpha}_i = 0, \quad \mu=1,\ldots,l.
\end{equation}
We denote by $U$ and $X$ the spaces of dependent
variables $u=(u^1,\ldots,u^q) \in \mathbb{R}^q$ and independent variables $x=(x^1,\ldots,x^p)\in \mathbb{R}^p$, respectively.  The functions $\Delta^{\mu i}_{\,\,\alpha}$ are assumed to be real valued functions on $U$ and are
components of the tensor products $\Delta^{\mu i}_{\,\,\alpha}\p_i \otimes du^{\alpha}$ on $X \times U$.  Here, we denote
the partial derivatives by $u^{\alpha}_i = \p_i u^{\alpha} \equiv \p u^{\alpha} / \p x^i$
and we adopt the convention that repeated indices are summed unless one of them is in a bracket.  For simplicity we
assume that all considered functions and manifolds are at least twice continuously differentiable in order to justify
our manipulations.  All our considerations have a local character.  For our purposes it suffices to search for solutions
defined on a neighborhood of the origin $x=0$.  In order to solve (\ref{EQ1-1}), we look for a map $f: X \rightarrow
J^1(X \times U)$ annihilating the contact 1-forms, i.e.
\begin{equation}
\label{EQ1-2}
f^{*}(du^{\alpha}-u^{\alpha}_i\,dx^i)=0.
\end{equation}
The image of $f$ is in a submanifold of the first jet space $J^1$ over $X$ given by (\ref{EQ1-1}) for which $J^1$ is
equipped with coordinates $x^i, u^{\alpha}, u^{\alpha}_i$.

This paper is organized as follows.  Section 2 contains a detailed
account of the construction of rank-1 solutions of PDEs (\ref{EQ1-1}).  In section 
3 we discuss the construction of rank-k solutions,
 using geometric and group invariant properties of the system (\ref{EQ1-1}). Section 4 deals with a
number of examples of hydrodynamic type systems which illustrate
the theoretical considerations. Several new classes of solutions in implicit and explicit form are obtained. Section 5
 contains a comparison of our results with the
generalized method of characteristics for multi-dimensional
systems of PDEs.

\section{The rank-1 solutions}

\text{}
\indent It is well known \cite{Cou1} that any hyperbolic system (\ref{EQ1-1}) admits rank-1 solutions
\begin{equation}
\label{EQ2-1}
u=f(r),\quad r(x,u)=\lambda_i(u)\,x^i,
\end{equation}
where $f=(f^{\alpha})$ are some functions of $r$ and a wave vector is a nonzero function
\begin{equation}
\label{EQ2-2}
\lambda(u)=\left(\lambda_1(u),\ldots,\lambda_p(u)\right)
\end{equation}
such that
\begin{equation}
\label{EQ2-3}
\ker{(\Delta^i\lambda_i)}\neq 0.
\end{equation}
Solution (\ref{EQ2-1}) is called a Riemann wave and the scalar function $r(x)$ is the Riemann invariant associated with the wave vector $\lambda$.

The function $f$ is a
solution of (\ref{EQ1-1}) if and only if the condition
\begin{equation}
\label{EQ2-4}
\left(\Delta^{\mu i}_{\,\,\alpha}(f) \lambda_i(f)\right){f'}^{\alpha}, \quad {f'}^{\alpha}=\frac{d f^{\alpha}}{d r}
\end{equation}
holds, i.e. if and only if $f'$ is an element of $\ker{(\Delta^i\,\lambda_i)}$.  Note that equation (\ref{EQ2-4}) is an
underdetermined system of the first order ordinary differential equations (ODEs) for $f$.  The image of a solution
(\ref{EQ2-1}) is a curve in $U$ space defined by the map $f:\mathbb{R} \rightarrow \mathbb{R}^q$ satisfying the set of
ODEs (\ref{EQ2-4}).  The extent to which expresion (\ref{EQ2-4}) constrains the function $f$ depends on the dimension of
$\ker{(\Delta^i\lambda_i)}$.  For example, if $\Delta^i\,\lambda_i=0$ then there is no constraint on the function $f$ at
all and no integration is involved.  The rank-1 solutions have the following common properties :\\
{\bf 1.} The Jacobian matrix is decomposable (in matrix notation)
\begin{equation}
\label{EQ2-5}
\p u = \left(1-\frac{\p f}{\p r}\,\frac{\p r}{\p u}\right)^{-1}\frac{\p f}{\p r}\,\lambda,
\end{equation}
or equivalently
\begin{equation}
\label{EQ2-6}
\p u = \frac{\p f}{\p r}\left(1-\frac{\p r}{\p u}\frac{\p f}{\p r}\right)^{-1}\,\lambda,
\end{equation}
where we have
\begin{equation}
\label{EQ2-7}
\begin{split}
&\p u = \left(u^{\alpha}_i\right) \in \mathbb{R}^{q \times p},\quad \frac{\p f}{\p r} = \left(\frac{\p f^{\alpha}}{\p
r}\right)\in \mathbb{R}^q,\\
&\frac{\p r}{\p u} = \left(\frac{\p r}{\p u^{\alpha}}\right)=\frac{\p \lambda_i}{\p u^{\alpha}} x^i \in \mathbb{R}^q,
\quad
\lambda = \left(\lambda_i\right) \in \mathbb{R}^p.
\end{split}
\end{equation}
This property follows directly from differentiation of (\ref{EQ2-1}).  The inverses $\left(1-\frac{\p f}{\p r}\frac{\p r}{\p u}\right)^{-1}$ or
$\left(1-\frac{\p r}{\p u}\frac{\p f}{\p r}\right)^{-1}$ are scalar functions and are defined, since
$\p r / \p u = 0$ at $x=0$.  From equations (\ref{EQ2-5}) or (\ref{EQ2-6}), it can be noted that $u(x)$ has rank at most
equal to 1. \\
{\bf 2.}  The graph of the rank-1 solution $\Gamma=\{x,u(x)\}$ is (locally) invariant under the linearly independent
vector fields
\begin{equation}
\label{EQ2-8}
X_a=\xi^i_a(u)\p_i, \quad a=1,\ldots,p-1
\end{equation}
acting on $X \times U$ space.  Here the vectors
\begin{equation}
\label{EQ2-9}
\xi_a(u)=\left(\xi^1_a(u),\ldots,\xi^p_a(u)\right)^T
\end{equation}
satisfy the orthogonality conditions
\begin{equation}
\label{EQ2-10}
\lambda_i\,\xi^i_a=0, \quad a=1,\ldots, p-1
\end{equation}
for a fixed wave vector $\lambda$ for which (\ref{EQ2-3}) holds.  The vector fields (\ref{EQ2-8}) span a Lie vector
module $g$ over functions on $U$ which constitutes an infinite-dimensional Abelian Lie algebra.  The algebra $g$
uniquely defines a
module $\Lambda$ (over the functions on $U$) of 1-forms $\lambda_i(u)\,dx^i$ annihilating all elements of $g$.  A basis
of $\Lambda$ is given by
\begin{equation}
\label{EQ2-11}
\lambda=\lambda_i(u)\, dx^i, \quad \xi^i_a\,\lambda_i=0
\end{equation}
for all indices $a=1,\ldots,p-1$.  The set $\{r=\lambda_i(u)x^i, u^1,\ldots,u^q\}$ is the complete set of invariants of
the vector fields (\ref{EQ2-8}).\\
{\bf 3.}  It should be noted that rescaling the wave vector $\lambda$ produces the same solution due to the homogeneity
of
the original
system (\ref{EQ1-1}).\\
{\bf 4.} Due to the orthogonality conditions (\ref{EQ2-10}), together with property (\ref{EQ2-5}) or (\ref{EQ2-6}), any
rank-1 solution is a solution of the overdetermined system of equations composed of system (\ref{EQ1-1}) and the 
differential constraints
\begin{equation}
\label{EQ2-12}
\xi^i_a(u)\,u^{\alpha}_i = 0, \quad a=1,\ldots,p-1.
\end{equation}
The side equations (\ref{EQ2-12}) mean that the characteristics of the vector fields (\ref{EQ2-8})
are equal to zero. \\
{\bf 5.} One can always find nontrivial solutions of (\ref{EQ2-4}) if (\ref{EQ1-1}) is an underdetermined system $(l<q)$
or if it is properly determined $(l=q)$ and hyperbolic.  Here, a weaker assumption can be imposed on the system
(\ref{EQ1-1}).  Namely, it is sufficient to require that eigenvalues of the matrix $(\Delta^i \lambda_i)$ are real
functions.\\

The method of construction of rank-1 solutions to (\ref{EQ1-1}) can be summarized as follows.  First, we seek a wave
vector
$\lambda=(\lambda_1,\ldots,\lambda_p)$ such that
\begin{equation}
\label{EQ2-14}
\rank{\left(\Delta^{\mu i}_{\,\,\alpha}\lambda_i\right)}<l.
\end{equation}
For each such choice of $\lambda_i$ we look for the solutions $\gamma^{\alpha}$ of the wave relations
\begin{equation}
\label{EQ2-15}
\left(\Delta^{\mu i}_{\,\,\alpha}\, \lambda_i\right)\gamma^{\alpha} = 0, \quad \mu=1,\ldots,l.
\end{equation}
Functions $f^{\alpha}(r)$ are required to satisfy the ODEs
\begin{equation}
\label{EQ2-16}
f'^{\alpha}(r)=\gamma^{\alpha}(f(r)).
\end{equation}
Alternatively, the system of equations (\ref{EQ2-4}) is linear in the variables $\lambda_i$.  Nonzero solutions
$\lambda_i$ exist if and only if
\begin{equation}
\label{EQ2-17}
\rank{\left({\Delta^{\mu}}^i_a\left(f(r)\right)f'^{\alpha}(r)\right)}<p.
\end{equation}
If (\ref{EQ2-17}) is satisfied for some function $f(r)$ then one can easily find $\lambda_i(r)$ satisfying equations (\ref{EQ2-4}).  Using
$u=f(r)$ one can define $\lambda_i(u)$ (not uniquely in general).  If $l<p$ then (\ref{EQ2-17}) is identically satisfied
for any function $f(r)$ and this approach does not require any integration.

\section{The rank-k solutions}

\label{sec:rank-k}
\text{}

This section is devoted to the construction of rank-k
solutions of a multi-dimensional system of PDEs (\ref{EQ1-1}).
These solutions may be considered as nonlinear superpositions of
rank-1 solutions.  

Suppose that we fix
$k$ linearly independent wave vectors $\lambda^1, \ldots, \lambda^k$, $1\leq k < p$ with  Riemann invariant
functions
\begin{equation}
\label{EQ3-1}
r^A(x,u)=\lambda^A_i(u) x^i, \quad A=1,\ldots,k.
\end{equation}
The equation
\begin{equation}
\label{EQ3-2}
u=f\left(r(x,u)\right), \quad r(x,u)=\left(r^1(x,u),\ldots,r^k(x,u)\right)
\end{equation}
then defines a unique function $u(x)$ on a neighborhood of $x=0$.  The Jacobian matrix of (\ref{EQ3-2}) is given by
\begin{equation}
\label{EQ3-3}
\p u = \left(\mathbb{I} - \frac{\p f}{\p r} \frac{\p r}{\p u} \right)^{-1}\frac{\p f}{\p r} \lambda,
\end{equation}
or equivalently
\begin{equation}
\label{EQ3-4}
\p u = \frac{\p f}{\p r} \left(\mathbb{I}-\frac{\p r}{\p u}\frac{\p f}{\p r}\right)^{-1}\lambda,
\end{equation}
where $f=(f^{\alpha})$, $f^{\alpha}$ are arbitrary functions of $r=(r^A)$ and
\begin{equation}
\label{EQ3-5}
\begin{split}
&\p u = (u^{\alpha}_i) \in \mathbb{R}^{q\times p}, \quad \frac{\p f}{\p r} = \left(\frac{\p f^{\alpha}}{\p r^A}\right)
\in \mathbb{R}^{q\times k},\\ &\lambda=\left(\lambda^A_i\right) \in \mathbb{R}^{k\times p},\quad
\frac{\p r}{\p u} = \left(\frac{\p r^A}{\p u^{\alpha}}\right) = \frac{\p \lambda^A_i}{\p u^{\alpha}}x^i \in
\mathbb{R}^{k \times q}.
\end{split}
\end{equation}
We assume here that the inverse matrices appearing in expressions (\ref{EQ3-3}) or (\ref{EQ3-4}), denoted by
\begin{equation}
\label{EQ3-6}
\Phi^1=\left(\mathbb{I}-\frac{\p f}{\p r}\frac{\p r}{\p u}\right)\in \mathbb{R}^{q \times q}, \quad \Phi^2=\left(\mathbb{I}-\frac{\p r}{\p u}\frac{\p f}{\p r}\right) \in \mathbb{R}^{k \times k}
\end{equation}
respectively, are invertible in some neighborhood of the origin
$x=0$.  This assumption excludes the gradient catastrophe
phenomenon for the function $u$. 

Note that the rank of the
Jacobian matrix (\ref{EQ3-3}) or (\ref{EQ3-4}) is at most equal to $k$.  Hence the image of the rank-k solution is a
k-dimensional submanifold $\mathcal{S}$ which lies in a submanifold of $J^1$.

If the set of vectors
\begin{equation}
\label{EQ3-7}
\xi_a(u)=\left(\xi^1_a(u),\ldots,\xi^p_a(u)\right)^T, \quad a=1,\ldots,p-k,
\end{equation}
satisfies the orthogonality conditions
\begin{equation}
\label{EQ3-8}
\lambda^A_i\,\xi^i_a =0
\end{equation}
for $A=1,\ldots,k$, $a=1,\ldots,p-k$ then by virtue of (\ref{EQ3-3}) or (\ref{EQ3-4}) we have
\begin{equation}
\label{EQ3-9} Q^{\alpha}_a(x,u^{(1)})\equiv
\xi^i_a\left(u\right)u^{\alpha}_i=0,\quad
a=1,\ldots,p-k,\quad \alpha=1,\ldots,q.
\end{equation}

Therefore rank-k solutions, given by (\ref{EQ3-2}), are
obtained from the overdetermined system (\ref{EQ1-1}) subjected to differential constraints (DCs) (\ref{EQ3-9})
\begin{equation}
\label{EQ3-10}
\Delta^{\mu i}_{\,\,\alpha}(u) u^{\alpha}_i=0,\quad
\xi^i_a(u) u^{\alpha}_i=0, \quad a=1,\ldots,p-k.
\end{equation}
Note that the conditions (\ref{EQ3-9}) are more general than the one required for the existence of Riemann k-wave
solutions (see expression (\ref{EQ5-1}) and discussion in Section 5). 

Let us note also that there are different approaches to the overdetermined system (\ref{EQ3-10}) employed in 
different versions of Riemann invariant method for multi-dimensional PDEs.  The essence of our approach lies 
in treating the problem from the point of view of the conditional symmetry method (for description see e.g. \cite{Olv1}).  
Below we proceed with the adaptation of this method for our purpose.

The graph of the rank-k solution $\Gamma=\{x,u(x)\}$ of (\ref{EQ3-10}) is invariant under the vector fields
\begin{equation}
\label{EQ3-11}
X_a=\xi^i_a(u)\p_i,\quad a=1,\ldots, p-k
\end{equation}
acting on $X \times U \subset \mathbb{R}^p \times \mathbb{R}^q$.  The functions $\{r^1,\ldots,r^k,u^1,\ldots,u^q\}$
constitute a complete set of invariants of the Abelian Lie algebra $\mathcal{A}$ generated by the vector fields (\ref{EQ3-11}).

In order to solve the overdetermined system (\ref{EQ3-10}) we subject it to several transformations, based on the set of invariants of $\mathcal{A}$, which simplify its structure considerably.  To achieve this simplification we choose an appropriate system of coordinates on $X \times U$ space which allows us to rectify the vector fields $X_a$, given by (\ref{EQ3-11}).  Next, we show how to find the invariance conditions in this system of coordinates which guarantee the existence of rank-k solutions in the form (\ref{EQ3-2}).

Let us assume that the $k$ by $k$ matrix
\begin{equation}
\label{EQ3-12}
\Pi=\left(\lambda^A_i\right), \quad 1 \leq A, i \leq k < p
\end{equation}
built from the components of the wave vectors $\lambda^A$ is invertible. Then the linearly independent vector fields
\begin{equation}
\label{EQ3-13}
\begin{split}
&X_{k+1}=\p_{k+1}-\sum_{A,j=1}^k \left(\Pi^{-1}\right)^j_A \lambda^A_{k+1} \p_j,\\
&\vdots\\
&X_p=\p_p - \sum_{A,j=1}^k \left(\Pi^{-1}\right)^j_A \lambda^A_p \p_j,
\end{split}
\end{equation}
have the required form (\ref{EQ3-11}) for which the orthogonality conditions (\ref{EQ3-8}) are satisfied.  The change of independent and dependent variables
\begin{equation}
\label{EQ3-14}
\bar{x}^1=r^1(x,u),\ldots,\bar{x}^k=r^k(x,u),\quad \bar{x}^{k+1}=x^{k+1},\ldots,\bar{x}^p=x^p, \bar{u}^1,\ldots,\bar{u}^q=u^q
\end{equation}
permits us to rectify the vector fields $X_a$ and get
\begin{equation}
\label{EQ3-15}
X_{k+1}=\p_{\bar{x}^{k+1}},\ldots, X_p=\p_{\bar{x}^p}.
\end{equation}
Note that a p-dimensional submanifold is transverse to the projection $(x,u) \rightarrow x$ at $x=0$ if and only if it is transverse to the projection $(\bar{x},\bar{u}) \rightarrow \bar{x}$ at $\bar{x}=0$.  The transverse p-dimensional submanifolds invariant under $X_{k+1},\ldots,X_p$ are defined by the implicit equation of the form
\begin{equation}
\label{EQ3-16}
\bar{u}=f(\bar{x}^1,\ldots,\bar{x}^k).
\end{equation}
Hence, expression (\ref{EQ3-16}) is the general integral of the invariance conditions
\begin{equation}
\label{EQ3-17}
\bar{u}_{\bar{x}^{k+1}}=0,\ldots, \bar{u}_{\bar{x}^p}=0.
\end{equation}
The system (\ref{EQ1-1}) is subjected to the invariance conditions (\ref{EQ3-17}) and, when written in terms of new coordinates $(\bar{x},\bar{u}) \in X \times U$, takes the form
\begin{equation}
\label{EQ3-18}
\Delta^{\mu}\left(\Phi^1\right)^{-1} \frac{\p \bar{u}}{\p \bar{x}} \lambda = 0, \quad, \bar{u}_{\bar{x}^{k+1}}=0,\ldots, \bar{u}_{\bar{x}^p}=0,
\end{equation}
or
\begin{equation}
\label{EQ3-19}
\Delta^{\mu}\frac{\p \bar{u}}{\p \bar{x}} \left(\Phi^2\right)^{-1} \lambda = 0,\quad, \bar{u}_{\bar{x}^{k+1}}=0,\ldots, \bar{u}_{\bar{x}^p}=0,
\end{equation}
where the matrices $\Phi^1$ and $\Phi^2$ are given by
\begin{equation}
\label{EQ3-20}
\left(\Phi^1\right)^A_i = \delta^A_i - \bar{u}^{\alpha}_i\frac{\p r^A}{\p \bar{u}^{\alpha}}, \quad \left(\Phi^2\right)^A_i = \delta^A_i - \frac{\p r^A}{\p \bar{u}^{\alpha}}\bar{u}^{\alpha}_i.
\end{equation}
The above considerations characterize geometrically the solutions of the overdetermined system (\ref{EQ3-10}) in the form (\ref{EQ3-2}).  Let us illustrate these considerations with some examples.

{\bf Example 1.} Let us assume that there exist $k$ independent relations of dependence for the matrices $\Delta^1,\ldots,\Delta^p$ such that the conditions
\begin{equation}
\label{EQ3-21}
\Delta^{\mu i}_{\,\,\alpha} \lambda^A_i = 0, \quad A=1,\ldots,k
\end{equation}
hold.  Suppose also that the original system (\ref{EQ1-1}) has the evolutionary form and each of the $q$ by $q$ matrices $A^1,\ldots,A^n$ is scalar, i.e.
\begin{equation}
\label{EQ3-22}
\Delta^0 = \mathbb{I},\quad  \Delta^{i\alpha}_{\beta} = a^i(u)\delta^{\alpha}_{\beta},\quad i=1,\ldots,n
\end{equation}
for some functions $a^1,\ldots,a^n$ defined on $U$, where $p=n+1$ and for convenience we denote the independent variables by $x=(t=x^0,x^1,\ldots,x^n) \in X$.  Then the system (\ref{EQ1-1}) is particularly simple and becomes
\begin{equation}
\label{EQ3-23}
u_t+a^1(u)u_1+\ldots+a^n(u)u_n=0.
\end{equation}
The corresponding wave vectors
\begin{equation}
\label{EQ3-24}
\begin{split}
&\lambda^1 = (-a^1(u),1,0,\ldots,0),\\
&\vdots\\
&\lambda^n = (-a^n(u),0,\ldots,0,1)
\end{split}
\end{equation}
are linearly independent and satisfy conditions (\ref{EQ3-21}).

A vector function $u(x,t)$ is a solution of (\ref{EQ3-23}) if and only if the vector field
\begin{equation*}
X=\p_t+a^i(u)\p_i
\end{equation*}
defined on $\mathbb{R}^{n+q+1}$ is tangent to the $(n+1)$-dimensional submanifold $\mathcal{S}=\{u=u(x,t)\} \subset \mathbb{R}^{n+q+1}$.  The solution is thus identified with the $(n+1)$-dimensional submanifold $\mathcal{S} \subset \mathbb{R}^{n+q+1}$ which is transverse to $\mathbb{R}^{n+q+1} \rightarrow \mathbb{R}^{n+1} : (x,t,u) \rightarrow (x,t)$ and is invariant under the vector field $X$.  The functions $\{r(x,t,u)=(r^1=x^1-a^1(u)t,\ldots, r^n=x^n-a^n(u)t),u^1,\ldots,u^q\}$ are invariants of $X$, such that $dr^1 \wedge \ldots \wedge dr^n \wedge du^1 \wedge \ldots \wedge du^q \neq 0$.  If we define $\bar{t}=t, \bar{u}=u$, then $(r,\bar{t},\bar{u})$ are coordinates on $\mathbb{R}^{n+q+1}$ and the vector field $X$ can be rectified
\begin{equation*}
X=\p_{\bar{t}}.
\end{equation*}
The general solution is
\begin{equation*}
\mathfrak{S}=\{F(r,\bar{u})=0\}
\end{equation*}
where $F:\mathbb{R}^{n+q} \rightarrow \mathbb{R}^q$ satisfies the condition
\begin{equation*}
\det{\left(\frac{\p F}{\p r} \frac{\p r}{\p \bar{u}} + \frac{\p F}{\p \bar{u}}\right)}\neq 0
\end{equation*}
but is otherwise arbitrary.  Note that it may be assumed that
\begin{equation*}
\frac{\p r}{\p u} (x_0,t_0,u_0)=0,
\end{equation*}
in which case the transversality condition is
\begin{equation*}
\det{\left(\frac{\p F}{\p \bar{u}}(x_0,t_0,u_0)\right)} \neq 0.
\end{equation*}
Hence the general solution of (\ref{EQ3-23}) near $(x_0,t_0,u_0)$ is
\begin{equation*}
\mathfrak{S}=\{\bar{u}=f(r)\},
\end{equation*}
where $f: \mathbb{R}^n \rightarrow \mathbb{R}^q$ is arbitrary. Thus the equation
\begin{equation}
\label{EQ3-25}
u=f(x^1-a^1(u)t,\ldots,x^n-a^n(u)t),
\end{equation}
defines a unique function $u(x,t)$ on a neighborhood of the point $(x_0,t_0,u_0)$ for any $f$.  Note that
\begin{equation*}
t=0, \quad u(x,0)=f(x^1,\ldots,x^n),
\end{equation*}
so the function $f$ is simply the Cauchy data on $\{t=0\}$.

{\bf Example 2.} Another interesting case to consider is when the matrix $\Phi^1$ (or $\Phi^2$) is a scalar matrix.  Then system (\ref{EQ3-18}) is equivalent to the quasilinear system in $k$ independent variables $\bar{x}^1,\ldots,\bar{x}^k$ and $q$ dependent variables $\bar{u}^1,\ldots,\bar{u}^q$.  So, we have
\begin{equation}
\label{EQ3-26}
B^A(\bar{u}) \bar{u}^{\alpha}_A=0,
\end{equation}
where
\begin{equation}
\label{EQ3-27}
B^A=\Delta^i \lambda^A_i.
\end{equation}
If $k \geq 2$ then $\Phi^1$ is a scalar if and only if
\begin{equation}
\label{EQ3-28}
\frac{\p r^1}{\p u}=0, \ldots, \frac{\p r^k}{\p u}=0
\end{equation}
and consequently, if and only if the vector fields $\lambda^1,\ldots, \lambda^k$ are constant wave vectors.

Finally, a more general situation occurs when the matrix $\Phi^1$ (or $\Phi^2$) satisfies the conditions
\begin{equation}
\label{EQ3-29}
\frac{\p \Phi^1}{\p \bar{x}^{k+1}}=0, \ldots, \frac{\p \Phi^1}{\p \bar{x}^p}=0.
\end{equation}
Then the system (\ref{EQ3-18}) is independent of variables $\bar{x}^{k+1},\ldots,\bar{x}^p$.  The conditions (\ref{EQ3-29}) hold if and only if
\begin{equation}
\label{EQ3-30}
\frac{\p^2 r}{\p u \p \bar{x}^{k+1}} = 0, \ldots, \frac{\p^2 r}{\p u \p \bar{x}^p} = 0.
\end{equation}
Using (\ref{EQ3-1}) and (\ref{EQ3-12}) we get
\begin{equation}
\label{EQ3-31}
\frac{\p \lambda^A_i}{\p u} = \sum_{l,B=1}^k \frac{\p \Pi^A_l}{\p u} \left(\Pi^{-1}\right)^l_B \lambda^B_i.
\end{equation}
Equation (\ref{EQ3-31}) can be rewritten in the simpler form
\begin{equation}
\label{EQ3-32}
\frac{\p}{\p u} \left(\sum_{B=1}^k \left(\Pi^{-1}\right)^l_B \lambda^B_i\right)=0,\quad 1 \leq l \leq k < i \leq p.
\end{equation}
Thus system (\ref{EQ3-18}) is independent of variables
$\bar{x}^{k+1},\ldots,\bar{x}^p$ if the $k$ by $p-k$ matrix
$\left(\lambda^B_i\right)$, $1 \leq B \leq k < i \leq p$ is equal
to the matrix $\Pi C$, where $C$ is a constant $k$ by $(p-k)$ matrix.
In this case (\ref{EQ3-18}) is a system not necessarily
quasilinear, in $k$ independent variables
$\bar{x}^1,\ldots,\bar{x}^k$ and $q$ dependent variables
$\bar{u}^1,\ldots,\bar{u}^q$. 

Let us now derive the neccesary and sufficient conditions for existence of solutions in the form (\ref{EQ3-2}) of the overdetermined system (\ref{EQ3-10}).
Substituting (\ref{EQ3-3}) or (\ref{EQ3-4}) into (\ref{EQ1-1}) yields
\begin{equation}
\label{EQ3-33}
\tr{\left[\Delta^{\mu}\left(\mathbb{I}-\frac{\p f}{\p r}\frac{\p r}{\p u}\right)^{-1}\frac{\p f}{\p r}
\lambda\right]}=0,
\end{equation}
or equivalently
\begin{equation}
\label{EQ3-34}
\tr{\left[\Delta^{\mu}\frac{\p f}{\p r}\left(\mathbb{I}-\frac{\p r}{\p u}\frac{\p f}{\p r}\right)^{-1}\lambda\right]}=0,
\end{equation}
respectively, where
\begin{equation}
\label{EQ3-35}
\Delta^{\mu} = \left(\Delta^{\mu i}_{\,\,\alpha}\right) \in \mathbb{R}^{p \times q}, \quad \mu=1,\ldots,l.
\end{equation}
Given the system of PDEs (\ref{EQ1-1}) (i.e. functions $\Delta^{\mu i}_{\,\,\alpha} (u)$) it follows that equations
(\ref{EQ3-33}) (or (\ref{EQ3-34})) are conditions on the functions $f^{\alpha}(r)$ and $\lambda^A_i(u)$ (or
$\xi^i_a(u)$).  Since $\p r / \p u$ depends explicitly on $x$ it may happen that these conditions have only trivial
solutions (i.e. $f$=const) for some values of $k$.  We discuss a set of conditions following from (\ref{EQ3-33}) or
(\ref{EQ3-34}) which allow the system (\ref{EQ3-10}) to possess the nontrivial rank-k solutions.

Let $g$ be a (p-k)-dimensional Lie vector module over $C^{\infty}(X \times U)$ with generators $X_a$ given by
(\ref{EQ3-11}).  Let $\Lambda$ be a k-dimensional module generated by $k<p$ linearly independent 1-forms
\begin{equation*}
\lambda^A=\lambda^A_i(u) dx^i, \quad A=1,\ldots,k
\end{equation*}
which are annihilated by $X_a \in g$.  It is assumed here that the vector fields $X_a$ and $\lambda^A$ are related by
the orthogonality conditions (\ref{EQ3-8}) and form a basis of $g$ and $\Lambda$, respectively.  For $k>1$, it is
always
possible to choose a basis $\lambda^A$ of the module $\Lambda$ of the form
\begin{equation}
\label{EQ3-36}
\lambda^A=dx^{i_A}+\lambda^A_{i_a} \, dx^{i_a},
\quad A=1,\ldots, k
\end{equation}
where $(i_A,i_a)$ is a permutation of $(1,\ldots,p)$.  Here we split the coordinates $x^i$ into $x^{i_A}$ and $x^{i_a}$. Then from (\ref{EQ3-1}) we obtain the relation
\begin{equation}
\label{EQ3-37}
\frac{\p r^A}{\p u^{\alpha}} = \frac{\p \lambda^A_{i_a}}{\p u^{\alpha}} x^{i_a}.
\end{equation}
Substituting (\ref{EQ3-37}) into equations (\ref{EQ3-33}) or (\ref{EQ3-34}) yields, respectively
\begin{equation}
\label{EQ3-38}
\tr{\left(\Delta^{\mu}(\mathbb{I}-Q_ax^{i_a})^{-1}\frac{\p f}{\p r}\lambda\right)}=0,
\end{equation}
or
\begin{equation}
\label{EQ3-39}
\tr{\left(\Delta^{\mu}\frac{\p f}{\p r} (\mathbb{I}-K_ax^{i_a})^{-1}\lambda\right)}=0,
\end{equation}
where we use the following notation
\begin{eqnarray}
\label{EQ3-40}
& &Q_a=\frac{\p f}{\p r} \eta_a \in \mathbb{R}^{q \times q}, \quad K_a=\eta_a \frac{\p f}{\p r} \in \mathbb{R}^{k \times
k},\\
\label{EQ3-41}
& &\eta_a=\left(\frac{\p \lambda^A_{i_a}}{\p u^{\alpha}}\right) \in \mathbb{R}^{k \times q}, \quad i_a=1,\ldots,p-1.
\end{eqnarray}
The functions $r^A$ and $x^{i_a}$ are all independent in the neighborhood of the origin $x=0$.  The functions $\Delta^{\mu}, \frac{\p
f}{\p r}, \lambda, Q_a$ and $K_a$ depend on $r$ only.  For these specific functions, equations (\ref{EQ3-38}) (or
(\ref{EQ3-39})) must be satisfied for all values of coordinates $x^{i_a}$.  In order to find appropriate conditions for $f(r)$
and $\lambda(u)$ let us notice that, according to the Cayley-Hamilton theorem, for any $n$ by $n$ invertible
matrice
M, $(M^{-1}\det{M})$ is a polynomial in $M$ of order $(n-1)$.  Hence, one can replace equation (\ref{EQ3-38}) by
\begin{equation}
\label{EQ3-42}
\tr{\left(\Delta^{\mu}\,Q\,\frac{\p f}{\p r}\,\lambda\right)}=0,
\end{equation}
where we introduce the following notation
\begin{equation*}
Q=(\mathbb{I}-Q_ax^{i_a})^{-1} \det{(\mathbb{I}-Q_ax^{i_a})}.
\end{equation*}
Taking equation (\ref{EQ3-42}) and all its $x^{i_a}$ derivatives (with $r$=const) at $x^{i_a}=0$, yields
\begin{eqnarray}
\label{EQ3-43}
\tr{\left(\Delta^{\mu}\frac{\p f}{\p r} \lambda\right)}=0,\\
\label{EQ3-44}
\tr{\left(\Delta^{\mu} Q_{(a_1}\ldots Q_{a_s)} \frac{\p f}{\p r} \lambda\right)}=0,
\end{eqnarray}
where $s=1,\ldots,q-1$ and $(a_1,\ldots,a_s)$ denotes symmetrization over all indices in the bracket.  A similar
procedure for equation (\ref{EQ3-39}) yields (\ref{EQ3-43}) and the trace condition
\begin{equation}
\label{EQ3-45}
\tr{\left(\Delta^{\mu}\,\frac{\p f}{\p r} K_{(a_1},\ldots,K_{a_s)} \lambda\right)}=0,
\end{equation}
where now $s=1,\ldots,k-1$.\\
Equation (\ref{EQ3-43}) represents an initial value condition on a surface in $X$ space given by $x^{i_a}=0$.  Equations
(\ref{EQ3-44}) (or (\ref{EQ3-45})) correspond to the preservation of (\ref{EQ3-43}) by flows represented by the vector
fields (\ref{EQ3-11}).  Note that $X_a$ can be put into the form
\begin{equation}
\label{EQ3-46}
X_a=\p_{i_a}-\lambda^A_{i_a}\p_A,\quad \xi^i_a\cdot \lambda^A_i = 0, \quad A=1,\ldots,k.
\end{equation}
By virtue of (\ref{EQ3-40}), (\ref{EQ3-41}), equations (\ref{EQ3-44}) or (\ref{EQ3-45}) take the unified form
\begin{equation}
\label{EQ3-47}
\tr{\left(\Delta^{\mu}\,\frac{\p f}{\p r}\eta_{(a_1}\frac{\p f}{\p r}\ldots \eta_{a_s)}\frac{\p f}{\p
r}\lambda\right)}=0,
\end{equation}
where either $\max{s}=q-1$ or $\max{s}=k-1$.

The vector fields $X_a$ and the Lie module $g$ spanned by the vector fields $X_1,\ldots, X_{p-k}$ are
called the conditional symmetries and the conditional symmetry module of (\ref{EQ1-1}), respectively if $X_a$ are Lie
point symmetries of the original system (\ref{EQ1-1}) supplemented by the DCs (\ref{EQ3-9}) \cite{Olv1}.  

Let us now associate the system (\ref{EQ1-1}) and the conditions (\ref{EQ3-9}) with the subvarieties of the solution spaces
\begin{equation*}
\mathcal{B}_{\Delta}=\{(x,u^{(1)}) : \Delta^{\mu i}_{\,\,\alpha}(u)\,u^{\alpha}_i = 0, \quad \mu=1,\ldots,l\},
\end{equation*}
and
\begin{equation*}
\mathcal{B}_Q=\{(x,u^{(1)}) : \xi^i_a(u)u^{\alpha}_i=0, \quad a=1,\ldots,p-k, \quad \alpha=1,\ldots,q\},
\end{equation*}
respectively.  We have the following.

\begin{theo}
A nondegenerate first order hyperbolic system of PDEs (\ref{EQ1-1}) admits a (p-k)-dimensional Lie vector module $g$ of conditional symmetries if and only if (p-k) linearly independent vector fields $X_1,\ldots,X_{p-k}$ satisfy the conditions (\ref{EQ3-43}) and (\ref{EQ3-47}) on some neighborhood of $(x_0,u_0)$ of $\mathcal{B}=\mathcal{B}_{\Delta} \cap \mathcal{B}_Q$.
\end{theo}

\begin{proof}
The vector fields $X_a$ constitute the conditional symmetry module $g$ for the system (\ref{EQ1-1}) if they are Lie point symmetries of the overdetermined ststem (\ref{EQ3-10}). This means that the first prolongation of $X_a$ has to be tangent to the system (\ref{EQ3-10}).  Hence $g$ is a conditional symmetry module of (\ref{EQ1-1}) if and only if the equations
\begin{equation}
\label{EQ3-48}
\mathrm{pr}^{(1)} X_a (\Delta^{\mu i}_{\,\,\alpha}(u)\,u^{\alpha}_i = 0, \quad \mathrm{pr}^{(1)} X_a \left(\xi^i_b(u)u^{\alpha}_i\right)=0, \quad a=1,\ldots,p-k
\end{equation}
are satisfied on $J^1$ whenever the equations (\ref{EQ3-10}) hold.
Now we show that if the conditions (\ref{EQ3-43}) and (\ref{EQ3-47}) are satisfied then the symmetry criterion
(\ref{EQ3-48}) is identically equal to zero.

In fact, applying the first prolongation of the vector fields $X_a$
\begin{equation*}
\mathrm{pr}^{(1)} X_a = X_a + \xi^i_{a,u^{\beta}}u^{\beta}_j u^{\alpha}_i \frac{\p}{\p u^{\alpha}_j}
\end{equation*}
to the original system (\ref{EQ1-1}) yields
\begin{equation}
\label{EQ3-49}
\mathrm{pr}^{(1)} X_a \left(\Delta^{\mu i}_{\,\,\alpha} u^{\alpha}_i\right) = \Delta^{\mu i}_{\,\,\alpha}
\xi^j_{a,u^{\beta}} u^{\beta}_iu^{\alpha}_j = 0,
\end{equation} 
whenever equations (\ref{EQ3-10}) hold.
On the other hand, carrying out the differentiations of (\ref{EQ3-8}) gives
\begin{equation}
\label{EQ3-50}
\xi^j_{a,u^{\beta}}\lambda^B_j=-\xi_a^j \lambda^B_{j,u^{\beta}}.
\end{equation}
Comparing (\ref{EQ3-49}) and (\ref{EQ3-50}) leads to
\begin{equation}
\label{EQ3-51}
{\Omega^{\mu}}^A_B \xi^j_a Z_A(\lambda^B_j)=0,
\end{equation}
where we introduce the following notation
\begin{equation}
\label{EQ3-52}
{\Omega^{\mu}}^A_B = \Delta^{\mu i}_{\,\,\alpha} Z^{\alpha}_B \lambda^A_i.
\end{equation}
Here the new vector fields $Z_B$ are defined on $U$
\begin{equation}
\label{EQ3-53}
Z_A=Z_A^{\alpha} \frac{\p}{\p u^{\alpha}} \in T_{u}U.
\end{equation}
It is convenient to write equation (\ref{EQ3-51}) in the equivalent form
\begin{equation}
\label{EQ3-54}
\tr{\left(\Delta^{\mu} Z \theta_a Z \lambda\right)}=0, \mu=1,\ldots,l
\end{equation}
where the following notation has been used
\begin{equation}
\label{EQ3-55}
\theta_a=\lambda^A_{i,u^{\beta}} \xi^i_a.
\end{equation}
The assumption that system (\ref{EQ1-1}) is hyperbolic implies that there exist the real-valued vector fields $\lambda^A$ and $\gamma_A$ defined on $U$ for which the wave relation
\begin{equation}
\label{EQ3-56}
\left(\Delta^{\mu\, i}_{\,\,\alpha}\lambda^A_i\right)\gamma^{\alpha}_{(A)}=0, \quad A=1,\ldots,k
\end{equation}
is satisfied and that the $U$ space is spanned by the linearly independent vector fields
\begin{equation}
\label{EQ3-57} \gamma_A=\gamma^{\alpha}_A\,\p_{u^{\alpha}} \in
T_{u}U.
\end{equation}
Hence, one can represent the vector fields $Z_A$ through the basis generated by the vector fields $\{\gamma_1,\ldots,\gamma_k\}$, i.e.
\begin{equation}
\label{EQ3-58}
Z_A=h^B_A\gamma_B.
\end{equation}
Using equations (\ref{EQ3-3}) and (\ref{EQ3-6}) we find the coefficients
\begin{equation*}
h^B_A=((\phi^1)^{-1})^B_A.
\end{equation*}
This means that the submanifold $\mathcal{S}$, given by (\ref{EQ3-2}), can be represented parametrically by
\begin{equation}
\label{EQ3-59}
\frac{\p f^{\alpha}}{\p r^A} = h^B_A \gamma^{\alpha}_B.
\end{equation}
On the other hand, comparing (\ref{EQ3-3}) and (\ref{EQ3-58}) gives
\begin{equation}
\label{EQ3-60}
u^{\alpha}_i=Z^{\alpha}_A \lambda^A_i.
\end{equation}
Applying the invariance criterion (\ref{EQ3-48}) to the side conditions (\ref{EQ3-9}) we obtain
\begin{equation}
\label{EQ3-61}
\mathrm{pr}^{(1)} X_a(Q^{\alpha}_b)=\xi^i_{[b}\xi^j_{a],u^{\beta}}u^{\beta}_iu^{\alpha}_j.
\end{equation}
The bracket $[a,b]$ denotes antisymmetrization with respect to the indices $a$ and $b$.  By virtue of equations (\ref{EQ3-50}) and (\ref{EQ3-60}), the right side of (\ref{EQ3-61}) is identically equal to zero.
Substituting (\ref{EQ3-58}) into equation (\ref{EQ3-54}) and taking into account equation (\ref{EQ3-36}) and (\ref{EQ3-59}) we obtain that for any value of $x \in X$ the resulting formulae coincide with equations (\ref{EQ3-43}) and (\ref{EQ3-47}).  Hence, the infinitesimal symmetry criterion (\ref{EQ3-48}) for the overdetermined system (\ref{EQ3-10}) is identically satisfied whenever conditions (\ref{EQ3-43}) and (\ref{EQ3-47}) hold.

The converse also holds.  The assumption that the system (\ref{EQ1-1}) is nondegenerate means that it is locally solvable and takes a maximal rank at every point $(x_0,u_0^{(1)}) \in \mathcal{B}_{\Delta}$.  Therefore \cite{Olv2} the infinitesimal symmetry criterion is a necessary and sufficient condition for the existence of symmetry group $G$ of the overdetermined system (\ref{EQ3-10}).  Since the vector fields $X_a$ form an Abelian distribution, it follows that the conditions (\ref{EQ3-43}) and (\ref{EQ3-47}) hold.  That ends the proof since the solutions of the original system (\ref{EQ1-1}) are invariant under the Lie algebra generated by $(p-k)$ vectors fields $X_1,\ldots,X_{p-k}$.
\end{proof}

Note that the set of solutions of the determining equations obtained by applying the symmetry criterion to the overdetermined system (\ref{EQ3-10}) is different than the set of solutions of the determining equations for the initial system (\ref{EQ1-1}).  Thus the system (\ref{EQ3-10}) other symmetries than the original system (\ref{EQ1-1}). So, new reductions for the system (\ref{EQ1-1}) can be constructed, since each solution of system (\ref{EQ3-10}) is a solution of system (\ref{EQ1-1}).

In our approach the construction of solutions of the original system (\ref{EQ1-1}) requires us to solve first the system
(\ref{EQ3-47}) for $\lambda^A_i$ as functions of $u^{\alpha}$ and then find $u=f(r)$ by solving (\ref{EQ3-43}).
Note that the functions $f^{*}(\lambda^A_i)$ are the functions $\lambda^A_i(f)$ pulled back to the surface $\mathcal{S}$.  The
$\lambda^A_i(f)$ then become functions of the parameters $r^1,\ldots,r^k$ on $\mathcal{S}$.  For simplicity of notation we denote
$f^{*}(\lambda^A_i)$ by $\lambda^A_i(r^1,\ldots,r^k)$.

The system composed of (\ref{EQ3-43}) and (\ref{EQ3-47}) is, in general, nonlinear.  So, we cannot expect to solve it in a closed form, except in some particular
cases.  But nevertheless, as we show in section 4, there are physically interesting examples for which
solutions of (\ref{EQ3-43}) and (\ref{EQ3-47}) lead to the new solutions of (\ref{EQ1-1}) which depend on some arbitrary functions.  These
particular solutions of (\ref{EQ3-43}) and (\ref{EQ3-47}) are obtained by expanding each function $\lambda^A_i$ into a
polynomial in the dependent variables $u^{\alpha}$ and requiring that the coefficients of the successive powers
of $u^{\alpha}$ vanish.  We then obtain a system of first order PDEs for the coefficients of the
polynomials.  Solving this system allows us to find some particular classes of solutions of the initial system
(\ref{EQ1-1}) which can be constructed by applying the symmetry reduction technique. 

\section{Examples of applications}

\label{sec:APPLICATIONS}
\text{}

We start with considering the case of rank-2 solutions of the system (\ref{EQ1-1}) with two dependent variables $(q=2)$.  Then (\ref{EQ3-47}) adopts the simplified form.
\begin{equation}
\label{EQ3-64}
\tr{\left(\Delta^{\mu}\frac{\p f}{\p r}\, \eta_a\,\frac{\p f}{\p r}\lambda\right)}=0.
\end{equation}
By virtue of (\ref{EQ3-43}), equation (\ref{EQ3-64}) can be transformed to
\begin{equation}
\label{EQ3-65}
\tr{\left[\Delta^{\mu}\frac{\p f}{\p r}\left(\eta_a\frac{\p f}{\p r} - \mathbb{I}\tr{\left(\eta_a\frac{\p f}{\p r}
\right)}\right)\lambda\right]}=0.
\end{equation}
Using the Cayley Hamilton identity, we get the relation
\begin{equation}
\label{EQ3-66}
AB-\mathbb{I}\,\tr{AB}=(B-\mathbb{I}\,\tr{B})(A-\mathbb{I}\,\tr{A})
\end{equation}
for any $2$ by $2$ matrices $A,B \in \mathbb{R}^{2 \times 2}$.  Now we can rewrite
(\ref{EQ3-65}) in the equivalent form
\begin{equation}
\label{EQ3-67}
-\tr{\left[\Delta^{\mu} \frac{\p f}{\p r} \left(\frac{\p f}{\p r}- \mathbb{I}\tr{\frac{\p f}{\p
r}}\right)(\eta_a-\mathbb{I}\tr{\eta_a})\lambda\right]}=0.
\end{equation}
So we have
\begin{equation}
\label{EQ3-68}
\det{\left(\frac{\p f}{\p r}\right)} \tr{[\Delta^{\mu}(\eta_a-\mathbb{I}\tr{\eta_a})\lambda]}=0.
\end{equation}
The rank-2 solutions require that the condition $\det{\p f/\p r}\neq 0$ be satisfied (otherwise $q=2$ can be
reduced to $q=1$).  As a consequence of this, we obtain the following condition
\begin{equation}
\label{EQ3-69}
\tr{[\Delta^{\mu}(\eta_a-\mathbb{I}\tr{\eta_a})\lambda]}=0,\quad \mu=1,\ldots,l,
\end{equation}
which coincides with the result obtained earlier for this specific case \cite{Gru4}.
One can look first for solutions $\lambda(u)$ of
(\ref{EQ3-69}) and then find $f(r)$ by solving (\ref{EQ3-43}).  Note that equations (\ref{EQ3-69}) form a system of
$l(p-2)$ equations for $2(p-2)$ functions $\lambda^A_{i_a}(u)$.  This indicates that they should have solutions (say,
for
generic systems) if (\ref{EQ1-1}) is not overdetermined.\\

{\bf Example 3.} We are looking for rank-2 solutions of the
(2+1) hydrodynamic type equations
\begin{equation}
\label{EQ4-1} u^i_t+u^ju^i_j + A^{ij}_{\,k} u^k_j=0, \quad
i,j,k=1,2
\end{equation}
where $A^{i}$ are some matrix functions of $u^1$ and $u^2$. Using the condition representing the tracelessness of the matrices
$\Delta^{1i}_{\,\alpha}u^{\alpha}_i$ and
$\Delta^{2i}_{\,\alpha}u^{\alpha}_i$, it is
convenient to rewrite the system (\ref{EQ4-1}) in the following form
\begin{equation}
\label{EQ4-2}
\begin{split}
&\tr{\left[ \left( \begin{array}{ccc} 1 & u^1+A^{11}_{\,1} & u^2+A^{12}_{\,1} \\
0
& A^{11}_{\,2} & A^{12}_{\,2}\end{array}\right) \left(\begin{array}{cc} u^1_t & u^2_t \\
u^1_x & u^2_x\\ u^1_y
& u^2_y \end{array}\right)\right]}=0, \\
&\tr{\left[\left(\begin{array}{ccc} 0 & u^1 & u^2 \\
1
& u^1+A^{21}_{\,2} & u^2+A^{22}_{\,2}\end{array}\right) \left(\begin{array}{cc} u^1_t & u^2_t \\
u^1_x & u^2_x\\ u^1_y & u^2_y
\end{array} \right)\right]}=0.
\end{split}
\end{equation}

Let $\mathcal{F}$ be a smooth orientable surface immersed in
$3$-dimensional Euclidean $(x,y,t) \in X$  space.  Suppose that
$\mathcal{F}$ can be written in the following parametric form
\begin{equation}
\label{EQ4-3} u=f(r^1,r^2)=(u^1(r^1,r^2),u^2(r^1,r^2)),
\end{equation}
such that the Jacobian matrix is different from zero
\begin{equation}
\label{EQ4-4} J=\det{\left(\frac{\p f^{\alpha}}{\p r^A}\right)} =
\det{\left(\begin{array}{cc} \p u^1/\p r^1 & \p u^1/\p r^2\\ \p
u^2/\p r^1 & \p u^2 /\p r^2 \end{array} \right)} \neq 0.
\end{equation}
Without loss of generality, it is possible to choose a basis
$\lambda^A$ of module $\Lambda$ of the form
\begin{equation}
\label{EQ4-5} \lambda^A_i = \left(\begin{array}{ccc} \lambda^1_1 &
\lambda^1_2 & \lambda^1_3 \\ \lambda^2_1 & \lambda^2_2 &
\lambda^2_3
\end{array} \right)^T = \left(\begin{array}{ccc} \varepsilon & a^1 & b^1\\ \varepsilon & a^2 & b^2 \end{array}\right)^T, \quad \varepsilon=\pm 1,
\end{equation}
where $a^A$ and $b^A$ are functions of $u^1$ and $u^2$ to be
determined.

The rank-2 solution can be constructed from the most general
solution of equations (\ref{EQ3-69}) for
$\lambda^A=(-1,a^A,b^A)$, $A=1,2$.
These equations lead to a system of four PDEs with four dependent variables
$a^A, b^A, A=1,2$ and two independent variables $u^1$ and $u^2$,
\begin{equation}
\label{EQ4-6}
\begin{split}
&-(A^{11}_2a^2+A^{12}_2b^2)\frac{\p a^1}{\p
u^1}+((u^1+A^{11}_1)a^2+(u^2+A^{12}_1)b^2-1)\frac{\p a^1}{\p
u^2}\\&\quad +(A^{11}_2a^1+A^{12}_2b^1)\frac{\p a^2}{\p
u^1}-((u^1+A^{11}_1)a^1+(u^2+A^{12}_1)b^1-1)\frac{\p a^2}{\p
u^2}=0,
\end{split}
\end{equation}
\begin{equation*}
\begin{split}
&-(A^{11}_2a^2+A^{12}_2b^2)\frac{\p b^1}{\p
u^1}+((u^1+A^{11}_1)a^2+(u^2+A^{12}_1)b^2-1)\frac{\p b^1}{\p
u^2}\\&\quad +(A^{11}_2a^1+A^{12}_2b^1)\frac{\p b^2}{\p
u^1}-((u^1+A^{11}_1)a^1+(u^2+A^{12}_1)b^1-1)\frac{\p b^2}{\p
u^2}=0,
\end{split}
\end{equation*}
\begin{equation*}
\begin{split}
&(1-(u^1+A^{21}_2)a^2-(u^2+A^{22}_2)b^2)\frac{\p a^1}{\p
u^1}+(A^{21_1}a^2+A^{22}_1b^2)\frac{\p a^1}{\p u^2}\\
& - (1-(u^1+A^{21}_2)a^1-(u^2+A^{22}_2)b^1)\frac{\p a^2}{\p u^1} -
(A^{21}_1a^1+A^{22}_1b^1)\frac{\p a^2}{\p u^2}=0,
\end{split}
\end{equation*}
\begin{equation*}
\begin{split}
&(1-(u^1+A^{21}_2)a^2-(u^2+A^{22}_2)b^2)\frac{\p b^1}{\p
u^1}+(A^{21_1}a^2+A^{22}_1b^2)\frac{\p b^1}{\p u^2}\\
& - (1-(u^1+A^{21}_2)a^1-(u^2+A^{22}_2)b^1)\frac{\p b^2}{\p u^1} -
(A^{21}_1a^1+A^{22}_1b^1)\frac{\p b^2}{\p u^2}=0.
\end{split}
\end{equation*}

Finally, a rank-2 solution of (\ref{EQ4-2}) is obtained
from the explicit parametrization of the surface $\mathcal{F}$ in
terms of the parameters $r^1$ and $r^2$, by solving 
equations (\ref{EQ3-43}) in which $\lambda^A$ adopt the form (\ref{EQ3-69})
\begin{equation}
\label{EQ4-7}
\begin{split}
&((u^1+A^{11}_1)a^1+(u^2+A^{12}_1)b^1-1)\frac{\p u^1}{\p
r^1}+((u^1+A^{11}_1)a^2+(u^2+A^{12}_1)b^2-1)\frac{\p u^1}{\p
r^2}\\
& \quad +(A^{11}_2a^1+A^{12}_2b^1)\frac{\p u^2}{\p r^1} +
(A^{11}_2a^2+A^{12}_2b^2)\frac{\p u^2}{\p r^2}=0,
\end{split}
\end{equation}
\begin{equation*}
\begin{split}
& (A^{21}_1a^1+A^{22}_1b^1)\frac{\p u^1}{\p r^1} +
(A^{21}_1a^2+A^{22}_1b^2)\frac{\p u^1}{\p r^2} + ((u^1+A^{21}_2)a^1+(u^2+A^{22}_2)b^1-1)\frac{\p u^2}{\p
r^1}\\
& \quad + ((u^1+A^{21}_2)a^2+(u^2+A^{22}_2)b^2-1)\frac{\p u^2}{\p
r^2}=0,
\end{split}
\end{equation*}
while the quantities $r^1$ and $r^2$ are implicitly defined as functions of $y,x,t$ by equation (\ref{EQ3-1}).\\

In the case when equation (\ref{EQ3-65}) does admit two linearly
independent vector fields $\lambda^A$ with $\varepsilon=-1$, there
exists a class of rank-2 solutions of equations (\ref{EQ4-6}) and
(\ref{EQ4-7}) invariant under the vector fields
\begin{equation}
\label{EQ4-8}
  X_1=\p_t+u^1\p_x, \quad X_2=\p_t+u^2\p_y.
\end{equation}
Following the procedure outlined in Section 3 we assume that the functions $f^1$ and $f^2$ appearing in equation
(\ref{EQ3-2}) are linear in $u^2$.  Then the invariance conditions
take the form
\begin{equation}
\label{EQ4-9}
  x-u^1t=g(u^1)+u^2h(u^1),\quad y-u^2t=a(u^1)+u^2b(u^1),
\end{equation}
where $a,b,g$ and $h$ are some functions of $u^1$.

One can show that if $h=0$, then the solution of the system
(\ref{EQ4-6}), (\ref{EQ4-7}) is defined implicitly by the
relations
\begin{equation}
\label{EQ4-10}
x-u^1t=g(u^1), \quad y-u^2t=a(u^1)+u^2g_{,u^1},
\end{equation}
where $a$ and $g$ are arbitrary functions of $u^1$. Note that in this case the functions $u^1$ and $u^2$ satisfy the following system of equations
\begin{equation}
\label{EQ4-11}
\begin{split}
&u^1_t+u^1u^1_x+u^2u^1_y + A^{11}_1(u^1_x-u^2_y)+A^{12}_1u^1_y=0,\\
&u^2_t+u^1u^2_x+u^2u^2_y + A^{21}_1(u^1_x-u^2_y)+A^{22}_1u^1_y=0,
\end{split}
\end{equation}
for any functions $A^{ij}_k$ of two variables $u^1$ and $u^2$.\\
If the function $h$ of $u^1$ does not vanish anywhere ($h\neq 0$) then the rank-2 solution is defined implicitly by equations (\ref{EQ4-9}) and satisfies the following system of PDEs
\begin{equation}
\label{EQ4-12}
\begin{split}
&u^1_t+u^1u^1_x+u^2u^1_y+A^{12}_2[u^2_y-u^1_x+l(u^1)u^2_x+m(u^1)u^1_y]=0,\\
&u^2_t+u^1u^2_x+u^2u^2_y+A^{22}_2[u^2_y-u^1_x+l(u^1)u^2_x+m(u^1)u^1_y]=0,
\end{split}
\end{equation}
where $A^{12}_2$ and $A^{22}_2$ are any functions of two variables $u^1$ and $u^2$.  Given the functions $l$ and $m$ of $u^1$, we can prescribe the functions $a$ and $b$ in expression (\ref{EQ4-9}) to find
\begin{equation}
\label{EQ4-13}
h=\int l\,b_{,u^1} du^1,\quad
g=\int [b-hm-a_{,u^1}]du^1.
\end{equation}

For instance, consider a rank-2 solution of equations (\ref{EQ4-6}) and (\ref{EQ4-7}) invariant under the vector fields
\begin{equation}
\label{EQ4-14}
X_1=\p_t+u^1\p_x, \quad X_2=\p_t-u^2\p_y
\end{equation}
with the wave vectors $\lambda^A$ which are the nonzero multiples of $\lambda^1=(u^1,-1,0)$ and $\lambda^2=(u^2,0,-1)$.  Then the solution is defined by the implicit relations
\begin{equation}
\label{EQ4-15}
\begin{split}
&x-u^1t=g(u^1),\\
&y+u^2t=h(u^1)+u^2g_{,u^1}.
\end{split}
\end{equation}
and satisfies the following system of equations
\begin{equation}
\label{EQ4-16}
\begin{split}
u^1_t+u^1u^1_x+u^2u^1_y+b(u^1,u^2)u^1_y=0,\\
u^2_t+u^1u^2_x+u^2u^2_y+c(u^1,u^2)u^1_y=0,
\end{split}
\end{equation}
where $b$ and $c$ are arbitrary functions of $u^1$ and $u^2$.

Thus, putting it all together, we see that the constructed solutions correspond to superpositions of two rank-1 solutions (i.e. simple waves) with local velocities $u^1$ and $u^2$, respectively.  According to \cite{Gru1}, if we choose the initial data ($t=0$) for the functions $u^1$ and $u^2$ sufficiently small and such that their first derivatives with respect to $x$ and $y$ will have compact and disjoint supports, then asymptotically there exists a finite time $t=T>0$ for which rank-2 solution decays in the exact way in two rank-1 solutions, being of the same type as in the initial moment.

{\bf Example 4.}  Consider the overdetermined hyperbolic system in ($2+1$) dimensions ($p=3$)
\begin{equation}
\label{EQ4-17}
\begin{split}
&\frac{\p \vec{u}}{\p t} + (\vec{u}\cdot \nabla)\vec{u} + ka \grad{a}=0\\
&\frac{\p a}{\p t}+(\vec{u}\cdot \nabla)a+k^{-1}a\dive{\vec{u}}=0,\\
&\frac{\p a}{\p x}=0, \quad \frac{\p a}{\p y}=0,
\end{split}
\end{equation}
describing the nonstationary isentropic flow of a compressible ideal fluid.  Here we use the following notations : $\vec{u}=(u^1,u^2)$ is the flow velocity, $a(t)=\left(\frac{\gamma p}{\rho}\right)^{1/2} \neq 0$ is the sound velocity which depends on $t$ only, $k=2(\gamma-1)^{-1}$ and $\gamma$ is the polytropic exponent.

The system (\ref{EQ4-17}) can be written in an equivalent form as
\begin{equation}
\label{EQ4-18}
\begin{split}
&\tr{\left[\left(\begin{array}{ccc} 1 & u^1 & u^2 \\ 0 & 0 & 0\\ 0 & ka & 0 \end{array} \right) \left(\begin{array}{ccc} u^1_t & u^2_t & a_t \\ u^1_x & u^2_x & 0 \\ u^1_y & u^2_y & 0 \end{array} \right)\right]}=0,\\
& \tr{\left[\left(\begin{array}{ccc} 0 & 0 & 0\\1 & u^1 & u^2 \\  0 & 0 & ka \end{array} \right) \left(\begin{array}{ccc} u^1_t & u^2_t & a_t \\ u^1_x & u^2_x & 0 \\ u^1_y & u^2_y & 0 \end{array} \right)\right]}=0,\\
& \tr{\left[\left(\begin{array}{ccc} 0 & k^{-1}a & 0\\0 & 0 & k^{-1}a \\  1 & u^1 & u^2 \end{array} \right) \left(\begin{array}{ccc} u^1_t & u^2_t & a_t \\ u^1_x & u^2_x & 0 \\ u^1_y & u^2_y & 0 \end{array} \right)\right]}=0.\\
\end{split}
\end{equation}
We are interested here in the rank-2 solutions of (\ref{EQ4-18}).
So, we require that conditions (\ref{EQ3-43}) and (\ref{EQ3-47}) be
satisfied.  This demand constitutes the necessary and
sufficient condition for the existence of a surface $\mathcal{F}$
written in a parametric form (\ref{EQ4-3}) for which equation
(\ref{EQ4-4}) holds.  In our case, $p=q=3$ and $k=2$, conditions (\ref{EQ3-43}) and
(\ref{EQ3-47}) become
\begin{equation}
\label{EQ4-19}
\tr{\left(\Delta^{\mu}\frac{\p f}{\p r}\lambda\right)}=0, \quad \mu=1,2,3,
\end{equation}
and
\begin{equation}
\label{EQ4-20}
\tr{\left(\Delta^{\mu}\frac{\p f}{\p r}\left(\eta_1\frac{\p f}{\p r}\eta_2+\eta_2\frac{\p f}{\p r}\eta_1\right)\frac{\p f}{\p r}\lambda\right)}=0,
\end{equation}
respectively.  Here, we assume the following basis for the wave vectors
\begin{equation}
\label{EQ4-21}
\lambda^A_i = \left(\begin{array}{ccc} \lambda^1_0 & \lambda^1_1 & \lambda^1_2 \\ \lambda^2_0 & \lambda^2_1 & \lambda^2_2\end{array}\right)^T = \left(\begin{array}{ccc} -1 & v^1 & w^1 \\ -1 & v^2 & w^2 \end{array} \right)^T,
\end{equation}
where $v^A$ and $w^A$ are some functions of $u^1$ and $u^2$ to be determined.  The $2$ by $3$ matrices $\eta_a$ and the $3$ by $2$ matrix $\p f/\p r$ take the form
\begin{equation}
\label{EQ4-22}
\begin{split}
&\eta_a=\frac{\p \lambda^A_{i_a}}{\p u^{\alpha}}=\left(\begin{array}{ccc} \p \lambda^1_{i_a}/\p u^1 & \p \lambda^1_{i_a} /\p u^2 & \p \lambda^1_{i_a} / \p a \\ \p \lambda^2_{i_a} /\p u^1 & \p \lambda^2_{i_a} / \p u^2 & \p \lambda^2_{i_a} / \p a\end{array}\right), \quad a=1,2\\
&\frac{\p f}{\p r} = \left(\begin{array}{cc} \p u^1/ \p r^1 & \p u^1 / \p r^2 \\ \p u^2 / \p r^1 & \p u^2 / \p r^2 \\ \p a/\p r^1 & \p a / \p r^2 \end{array}\right).
\end{split}
\end{equation}
Equations (\ref{EQ4-19}) lead to the following differential conditions
\begin{equation}
\label{EQ4-23}
\begin{split}
&\frac{\p u^1}{\p r^1} + \frac{\p u^1}{\p r^2}+(u^1-kav^2)\frac{\p u^2}{\p r^1} + (u^1-kaw^2)\frac{\p u^2}{\p r^2}+u^2(\frac{\p a}{\p r^1}+\frac{\p a}{\p r^2})=0,\\
&v^1\frac{\p u^1}{\p r^1} + w^1\frac{\p u^1}{\p r^2}+u^1(v^1\frac{\p u^2}{\p r^1} + w^1\frac{\p u^2}{\p r^2}) + (u^2v^1+kav^2)\frac{\p a}{\p r^1} \\& \qquad+ (u^2w^1+kaw^2)\frac{\p a}{\p r^2}=0,\\
&k(v^2\frac{\p u^1}{\p r^1}+w^2\frac{\p u^1}{\p r^2})-(a-kv^2u^1)\frac{\p u^2}{\p r^1}\\& \qquad-(a-w^2ku^1)\frac{\p u^2}{\p r^2}+(av^1+kv^2u^2)\frac{\p a}{\p r^1}+(aw^1+kw^2u^2)\frac{\p a}{\p r^2}=0.
\end{split}
\end{equation}
Assuming that we have found $v^A$ and $w^A$ as functions of $u^1$
and $u^2$, we have to solve (\ref{EQ4-20}) for the unknown
functions $u^1$ and $u^2$ in terms of $r^1$ and $r^2$.  The resulting expressions in the equations
(\ref{EQ4-20}) are rather complicated, hence we omit them here.
%\begin{equation}
%\label{EQ4-24}
%\end{equation}
Various rank-2 solutions are determined by a specification
of functions $v^A$ and $w^A$ in terms of $u^1$ and $u^2$.  By way of
illustration we show how to obtain a solution which depends on one
arbitrary function of two variables.

Let us suppose that we are interested in the rank-2 solutions
invariant under the vector fields
\begin{equation}
\label{EQ4-24} X_1=\p_t+u^1\p_x, \quad X_2=\p_t+u^2\p_y.
\end{equation}
So, the functions $r^1=x-u^1t$ and $r^2=y-u^2t$ are the Riemann invariants
of these vector fields.  Under this assumption, equations
(\ref{EQ4-19}) and (\ref{EQ4-20}) can be easily solved to obtain the Jacobian matrix
\begin{equation}
\label{EQ4-25}
J=\frac{\p(u^1,u^2)}{\p(r^1,r^2)} \neq 0
\end{equation}
which has the characteristic polynomial with
constant coefficients.  This means that the trace and determinant
of $J$ are constant,
\begin{equation}
\label{EQ4-26}
\begin{split}
(i)&\quad u^1_{r^1}+u^2_{r^2}=2C_1,\\
(ii)&\quad u^1_{r^1}u^2_{r^2}-u^1_{r^2}u^2_{r^1}=C_2.
\end{split}
\quad C_1,C_2 \in \mathbb{R}
\end{equation}
The trace condition (\ref{EQ4-26}(i)) implies that
there exists a function $h$ of $r^1$ and $r^2$ such that the
conditions
\begin{equation}
\label{EQ4-27} u^1=C_1r^1+h_{r^2}, \quad u^2=C_1r^2-h_{r^1},
\end{equation}
hold.  The determinant condition (\ref{EQ4-26}(ii)) requires that
the function $h(r^1,r^2)$ satisfies the Monge-Amp{\`e}re equation
\begin{equation}
\label{EQ4-28} h_{r^1r^1}h_{r^2r^2}-h_{r^1r^2}^2=C, \quad C \in
\mathbb{R}.
\end{equation}
Hence, the general integral of the system (\ref{EQ4-17}) has the
implicit form defined by the relations between the variables
$t,x,y,u^1$ and $u^2$
\begin{equation}
\label{EQ4-29}
\begin{split}
  &u^1=C_1(x-u^1t)+\frac{\p h}{\p r^2}(x-u^1t,y-u^2t),\\
  &u^2=C_1(y-u^2t)+\frac{\p h}{\p r^1}(x-u^1t,y-u^2t),\\
  &a=a_0\left((1+C_1t)^2+Ct^2\right)^{-1/k}, \quad a_0 \in
  \mathbb{R}
\end{split}
\end{equation}
where the function $h$ obeys (\ref{EQ4-28}).  

Note that the
Gaussian curvature $K$ expressed in curvilinear coordinates
$(t,r^1,r^2) \in \mathbb{R}^3$ of the surface
$\mathcal{S}=\{t=h(r^1,r^2)\}$ is not constant and is given by
\begin{equation}
  \label{EQ4-30}
  K(r^1,r^2)=\frac{C}{1+h_{r^1}^2+h_{r^2}^2}.
\end{equation}
For example, a particular nontrivial class of solution of
(\ref{EQ4-17}) can be obtained if we assume that $C=0$.  In this
case the general solution of (\ref{EQ4-17}) depends on three
parameters, $a_0,C_1,m \in \mathbb{R}$ and takes the form
\begin{equation}
  \label{EQ4-31}
  \begin{split}
    &u^1=C_1(x-u^1t)+(1-m)\left(\frac{x-u^1t}{y-u^2t}\right)^m,\\
    &u^2=C_1(y-u^2t)-m\left(\frac{y-u^2t}{x-u^1t}\right)^{1-m},\\
    &a(t)=\frac{a_0}{(1+C_1t)^{2/k}}.
  \end{split}
\end{equation}
Note that if $C=0$ and $C_1=0$ then the Jacobian matrix $J$ is nilpotent and the divergence of the vector $\vec{u}$ is equal to zero.  Then
the expression
\begin{equation}
  \label{EQ4-32}
  \begin{split}
    &u^1=(1-m)\left(\frac{x-u^1t}{y-u^2t}\right)^m, \quad a=a_0,\\
    &u^2=-m\left(\frac{y-u^2t}{x-u^1t}\right)^{1-m},
  \end{split}
\end{equation}
defines a solution $\vec{u}=(u^1,u^2)$ to incompressible Euler equations
\begin{equation}
  \label{EQ4-33}
  \vec{u}_t+(\vec{u}\cdot \nabla)\vec{u}=0, \quad
  \dive{\vec{u}}=0,\quad a=a_0.
\end{equation}
Note that for $m=2$, the explicit form of (\ref{EQ4-32}) is
\begin{equation}
\begin{split}
\label{EQ4-33a}
&u^1 = \frac{-y^2-2tx \pm \sqrt{y^2+4tx}}{2t^2},\\
&u^2 = \frac {y \mp \sqrt{{y}^{2}+4tx}}{t}.
\end{split}
\end{equation}
{\bf Example 5.} Now let us consider a more general case of to the autonomous system (\ref{EQ4-17}) in $p=n+1$ independent $(t,x^i) \in X$ and $q=n+1$ dependent $(a,u^i) \in U$ variables. We look for the rank-k solutions, when $k=n$.  The change of variables in the system (\ref{EQ4-17}) under the point transformation
\begin{equation}
\label{EQ4-34}
\bar{t}=t,\quad \bar{x}^1=x^1-u^1t,\ldots,\bar{x}^n=x^n-u^nt, \quad \bar{a}=a,\quad \bar{u}=u
\end{equation}
leads to the following system
\begin{equation}
\label{EQ4-35}
\begin{split}
&\frac{D\bar{u}}{D\bar{t}}=0,\\
&\frac{D\bar{a}}{D\bar{x}}=0,\quad \frac{D\bar{a}}{D\bar{t}}+k^{-1}\bar{a}\tr{\left(B^{-1}\frac{D\bar{u}}{D\bar{x}}\right)}=0, \quad a\neq 0
\end{split}
\end{equation}
where the total derivatives are denoted by
\begin{equation}
\label{EQ4-36}
\frac{D}{D\bar{t}}=\frac{\p}{\p t}+\bar{u}^i_{\bar{t}}\frac{\p}{\p \bar{u}^i}, \quad \frac{D}{D\bar{x}^j}=\frac{\p}{\p \bar{x}^j}+\bar{u}^i_{\bar{x}^j}\frac{\p}{\p \bar{u}^i}, \quad j=1,\ldots,n
\end{equation}
and the $n$ by $n$ nonsigular matrix $B$ has the form
\begin{equation}
\label{EQ4-37}
B=\mathbb{I}+t\frac{\p \bar{u}}{\p \bar{x}}.
\end{equation}
The general solution of the first equation in (\ref{EQ4-35}) is
\begin{equation}
\label{EQ4-38}
\bar{u}=f(\bar{x}),\quad \bar{x}=(\bar{x}^1,\ldots,\bar{x}^n)
\end{equation}
for some function $f:\mathbb{R}^n \rightarrow \mathbb{R}^n$.  The second equation in (\ref{EQ4-35}) can be written in an equivalent form
\begin{equation}
\label{EQ4-39}
\frac{\p}{\p \bar{t}}\left(\ln{|\bar{a}(t)|^k}\right)+\tr{\left[(\mathbb{I}+\bar{t}Df(\bar{x}))^{-1}Df(\bar{x})\right]}=0,
\end{equation}
where the Jacobian matrix is denoted by
\begin{equation}
\label{EQ4-40}
Df(\bar{x})=\frac{\p f}{\p \bar{x}}(\bar{x}).
\end{equation}
Differentiation of equation (\ref{EQ4-39}) with respect to $\bar{x}$ yields
\begin{equation}
\label{EQ4-41}
\frac{\p^2}{\p \bar{x}\p\bar{t}}\left(\ln{\det{\left(\mathbb{I}+\bar{t}Df(\bar{x})\right)}}\right)=0
\end{equation}
with general solution
\begin{equation}
\label{EQ4-42}
\det{\left(\mathbb{I}+\bar{t}Df(\bar{x})\right)}=\alpha(\bar{x})\beta(\bar{t})
\end{equation}
for some functions $\alpha:\mathbb{R}^n \rightarrow \mathbb{R}$ and $\beta : \mathbb{R}\rightarrow \mathbb{R}$.  Evaluating (\ref{EQ4-42}) at the initial data $t=0$ implies $\alpha(\bar{x})=\beta(0)^{-1}$. Therefore
\begin{equation}
\label{EQ4-43}
\det{\left(\mathbb{I}+\bar{t}Df(\bar{x})\right)}=\frac{\beta(\bar{t})}{\beta(0)}.
\end{equation}
So, we have
\begin{equation}
\label{EQ4-44}
\frac{\p}{\p x} \det{\left(\mathbb{I}+\bar{t}Df(\bar{x})\right)}=0.
\end{equation}
Now, let us write the determinant in the form of the characteristic polynomial
\begin{equation}
\label{EQ4-45}
\det{\left(\mathbb{I}+\bar{t}Df(\bar{x})\right)}=\bar{t}^nP_n(\varepsilon,\bar{x}), \quad \varepsilon=\frac{1}{\bar{t}}
\end{equation}
where
\begin{equation}
\label{EQ4-46}
\det{\left(\varepsilon \mathbb{I} + Df(\bar{x})\right)}=\varepsilon^n + p_{n-1}(\bar{x})\varepsilon^{n-1}+\ldots+p_1(\bar{x})\varepsilon + p_0(\bar{x}).
\end{equation}
Equation (\ref{EQ4-44}) holds if and only if the coefficients of the characteristic polynomial $p_0,\ldots,p_{n-1}$ are constants.  So, equation (\ref{EQ4-39}) implies that
\begin{equation*}
\frac{\p}{\p \bar{t}}\ln{|\bar{a}(\bar{t})|^k} + \frac{\p}{\p \bar{t}}\ln{|\det{\left(\mathbb{I}+\bar{t}Df(\bar{x})\right)}|}=0.
\end{equation*}
Then we have,

\begin{equation}
\label{EQ4-47}
\frac{\p}{\p \bar{t}}\left(|\bar{a}(\bar{t})|^k\det{\left(\mathbb{I}+\bar{t}Df(\bar{x})\right)}\right)=0.
\end{equation}
Solving equation (\ref{EQ4-47}) we obtain
\begin{equation}
\label{EQ4-48}
\bar{a}(\bar{t}) = \gamma\left(\det{\left(\mathbb{I}+\bar{t}Df(\bar{x})\right)}\right)^{-1/k}, \quad 0 \neq \gamma \in  \mathbb{R}.
\end{equation}
Thus, the general solution of system (\ref{EQ4-17}) is
\begin{equation}
\label{EQ4-49} u=f(\bar{x}), \quad
a(\bar{t})=\gamma\left[1+p_{n-1}\bar{t}+\ldots+p_0\bar{t}^n\right]^{-1/k},
\end{equation}
for any differentiable function $f: \mathbb{R}^n \rightarrow \mathbb{R}^n$ and takes the form of a constant characteristic polynomial on the Cauchy data $t=0$
\begin{equation}
\label{EQ4-50}
P_n(\varepsilon) = \varepsilon^n+ p_{n-1}\varepsilon^{n-1} + \ldots + p_1\varepsilon + p_0.
\end{equation}
Note that the function $a$ is constant if and only if
\begin{equation}
\label{EQ4-51}
P_n(\varepsilon) = \varepsilon^n.
\end{equation}
This fact holds if and only if the Jacobian matrix $Df(\bar{x})$ is nilpotent.

Note that in the particular case when $p=2$, the general explicit solution of (\ref{EQ4-17}) is given by
\begin{equation}
\label{EQ4-52}
u(x,t)=(\beta+\alpha x)(1+\alpha t)^{-1},\quad a(t)=\gamma(1+\alpha t)^{-1/k}, \quad \alpha,\beta,\gamma \in \mathbb{R}.
\end{equation}
An extension of this solution to the (n+1)-dimensional space $X$ is as follows
\begin{equation}
\label{EQ4-53}
u(x,t)=(\mathbb{I}+t\alpha)^{-1}(\beta+\alpha x), \quad a(t)=\gamma(\det{(\mathbb{I}+\alpha t}))^{-1/k},
\end{equation}
where $\beta$ is any constant n-component vector and $\alpha$ is any $n$ by $n$ constant matrix.  In this case the Jacobian matrix is constant
\begin{equation}
\label{EQ4-54}
Df(\bar{x})=\alpha
\end{equation}
for any $\bar{x} \in \mathbb{R}^n$.

Finally, a similar computation can be performed for the case in which the vector function $\vec{u}=(u^1,u^2,u^3)$ satisfies the overdetermined system (\ref{EQ4-33}).  In the above notation, an invariant solution under the vector fields
\begin{equation}
\label{EQ4-55}
X_a=\p_t+u^a\p_{(a)}, \quad a=1,2,3
\end{equation}
is given by $\bar{u}=f(\bar{x})$ and is a divergence free solution
\begin{equation}
\label{EQ4-56}
\dive{\vec{u}}=0
\end{equation}
if and only if the trace condition
\begin{equation}
\label{EQ4-57}
\tr{\left(B^{-1}\frac{\p \bar{u}}{\p \bar{x}}\right)}=0, \quad B=\mathbb{I}+t\frac{D\bar{u}}{D\bar{x}}(\bar{x})
\end{equation}
holds.  But
\begin{equation}
\label{EQ4-58}
\frac{D\vec{u}}{D\bar{x}}=\frac{\p B}{\p t}.
\end{equation}
Therefore $\dive{\vec{u}}=0$ if and only if
\begin{equation}
\label{EQ4-59}
\tr{\left(B^{-1}\frac{\p B}{\p t}\right)}=0,
\end{equation}
or equivalently, if and only if
\begin{equation}
\label{EQ4-60}
\frac{\p}{\p t} \left(\det{B}\right)=0.
\end{equation}
This means that the Jacobian matrix $Df(\bar{x})$ has to be a nilpotent one and takes the form
\begin{equation}
\label{EQ4-61} Df(\bar{x}) = \left( \begin{array}{ccc}
0 & f^1_{\bar{x}^2} & f^1_{\bar{x}^3} \\
0 & f^2_s & -f^2_s\\
0 & f^2_s & -f^2_s
\end{array} \right),
\end{equation}
where $f^1$ is an arbitrary function of two variables $\bar{x}^2$ and $\bar{x}^3$ and $f^2$ is an arbitrary function of one variable $s=\bar{x}^2-\bar{x}^3$.
Note that if $f^1_{\bar{x}^2} \neq f^1_{\bar{x}^3}$ then the Jacobian matrix $Df(\bar{x})$ has rank $2$ (otherwise $f^1$ is any function of $s$ and $Df(\bar{x})$ has rank $1$).  As a consequence, the matrix $B$ has the form
\begin{equation}
\label{EQ4-62} B=\left( \begin{array}{ccc}
1 & tf^1_{\bar{x}^2} & tf^1_{\bar{x}^3}\\
0 & 1+tf^2_s & -tf^2_s\\
0 & tf^2_s & 1-tf^2_s
         \end{array}\right), \quad \det{B}=1.
\end{equation}
So, the condition (\ref{EQ4-60}) is identically satisfied.  Thus, the general solution of system (\ref{EQ4-33}) is implicitly defined by the equations
\begin{equation}
\label{EQ4-63}
u^1=f^1\big(x^2-tf^2(x^2-x^3),x^3-tf^2(x^2-x^3)\big), \quad
u^2=u^3=f^2(x^2-x^3), \quad a=a_0,
\end{equation}
where the functions $f^1$ and $f^2$ are arbitrary functions of their
arguments.  Equations (\ref{EQ4-63}) define a rank-2 solution but,
according to the formula for the corresponding principle \cite{Gru1}, it
is not a double Riemann wave.

Obviously, other choices of the wave vectors $\lambda^A$ (and the related
vector fields $X_a$) lead to different classes of solutions.  The
problem of the classification of these solutions remains still
open but some results are known (see e.g. the functorial
properties of systems of equations determining Riemann invariants
\cite{Gru3}).

\section{Conclusions}

\label{sec:CONC}
In this paper we have developed a new method which serves as a tool for constructing rank-k solutions of
multi-dimensional hyperbolic
systems including Riemann waves and their superpositions.  The most significant characteristic of this approach is that
it allows us to construct regular algorithms for finding solutions written in terms of Riemann invariants.  Moreover,
this approach does not refer to any additional considerations, but proceeds directly from the given system of PDEs.

Riemann waves and their superposition described by
multi-dimensional hyperbolic systems have been studied so far only
in the context of the generalized method of characteristics (GMC) \cite{Bur3,Gru1,Per2}.
The essence of this method can be summarized as follows.  It
requires the supplementation of the original system of PDEs
(\ref{EQ1-1}) with additional differential constraints for which
all first derivatives are decomposable in the following form
\begin{equation}
\label{EQ5-1}
\frac{\p u^{\alpha}}{\p x^i} (x) = \sum_{A=1}^k \xi^A(x) \gamma^{\alpha}_A (u) \lambda^A_i(u),
\end{equation}
where
\begin{equation}
\label{EQ5-2}
\begin{split}
&\left(\Delta^{\mu i}_{\,\,\alpha}(u)\lambda^A_i\right) \gamma^{\alpha}_{(A)} = 0, \quad A=1,\ldots, k\\
&\rank{\left(\Delta^{\mu i}_{\,\,\alpha}(u) \lambda^A_i\right)} < l.
\end{split}
\end{equation}
Here, $\xi^A \neq 0$ are treated as arbitrary scalar functions of
$x$ and we assume that the vector fields
$\{\gamma_1,\ldots,\gamma_k\}$ are locally linearly independent.
The necessary and sufficient conditions for the existence of
k-wave solutoins (when $k>1$) of the system (\ref{EQ5-1}) in terms
of Riemann invariants impose some additional differential conditions on
the wave vectors $\lambda^A$ and the corresponding vector fields
$\gamma_A$, namely \cite{Per2}
\begin{equation}
\label{EQ5-3}
\begin{split}
&[\gamma_A,\gamma_B] \in \mathrm{span} \{\gamma_A,\gamma_B\},\\
&\mathcal{L}_{\gamma_{\beta}} \lambda^A \in \mathrm{span} \{\lambda^A,\lambda^B\}, \quad \forall A \neq B=1,\ldots,k,
\end{split}
\end{equation}
where $[\gamma_A,\gamma_B]$ denotes the commutator of the vector fields $\gamma_A$ and $\gamma_B$,
$\mathcal{L}_{\gamma_B}$ denotes the Lie derivatives along the vector fields $\gamma_B$.

Due to the homogeneity of the wave relation (\ref{EQ5-2}) we can choose, without loss of generality, a holonomic system
for the fields $\{\gamma_1,\ldots,\gamma_k\}$ by requiring a proper length for each vector $\gamma_A$ such that
\begin{equation}
\label{EQ5-4}
[\gamma_A,\gamma_B]=0, \quad \forall A \neq B = 1, \ldots, k.
\end{equation}
It determines a k-dimensional submanifold $\mathcal{S} \subset U$ obtained by solving the system of PDEs
\begin{equation}
\label{EQ5-5}
\frac{\p f^{\alpha}}{\p r^A} = \gamma^{\alpha}_A(f^1,\ldots,f^k)
\end{equation}
with solution $\pi: F \rightarrow U$ defined by
\begin{equation}
\label{EQ5-6}
\pi : (r^1,\ldots,r^k) \rightarrow \left(f^1(r^1,\ldots,r^k),\ldots,f^q(r^1,\ldots,r^k)\right).
\end{equation}
The wave vectors $\lambda^A$ are pulled back to the submanifold $\mathcal{S}$ and then $\lambda^A$ become functions of
the
parameters $r^1,\ldots,r^k$. Consequently, the requirements (\ref{EQ5-1}) and (\ref{EQ5-3}) take the form
\begin{eqnarray}
\label{EQ5-7}
& &\frac{\p r^A}{\p x^i}(x) = \xi^A(x)\lambda^A_i(r^1,\ldots,r^k),\\
\label{EQ5-8}
& &\frac{\p \lambda^A}{\p r^B} \in \mathrm{span}\{\lambda^A,\lambda^B\}, \quad \forall A \neq B=1,\ldots,k
\end{eqnarray}
respectively.  It has been shown \cite{Per2} that the conditions (\ref{EQ5-5}) and (\ref{EQ5-8}) ensure that the set of solutions
of system (\ref{EQ1-1}) subjected to (\ref{EQ5-1}), depends on $k$ arbitrary functions of one variable.  It has
also been
proved \cite{Per1} that all solutions, i.e. the general integral of the system (\ref{EQ5-7}) under conditions (\ref{EQ5-8})
can be obtained by solving, with respect to the variables $r^1,\ldots,r^k$, the system in implicit form
\begin{equation}
\label{EQ5-9}
\lambda^A_i(r^1,\ldots,r^k)x^i=\psi^A(r^1,\ldots,r^k),
\end{equation}
where $\psi^A$ are arbitrary functionnally independent differentiable functions of $k$ variables $r^1,\ldots,r^k$.  Note
that solutions of (\ref{EQ5-7}) are constant on $(p-k)$-dimensional hyperplanes perpendicular to wave vectors
$\lambda^A$ satisfying conditions (\ref{EQ5-8}).

As one can observe, both methods discussed here exploit the invariance properties of the initial system of equations
(\ref{EQ1-1}).  In the GMC, they have the purely geometric character for which a form of solution is postulated by
subjecting the original system (\ref{EQ1-1}) to the side conditions (\ref{EQ5-1}).  In contrast, in the case of the
approach proposed here we augment the system (\ref{EQ1-1}) by differential constraints (\ref{EQ3-9}).

There are, however, at least two basic differences between the GMC and our proposed approach.  Riemann multiple waves
defined from (\ref{EQ5-1}), (\ref{EQ5-5}) and (\ref{EQ5-8}) constitute a more limited class of solutions than the rank-k
solutions postulated by our approach.  This difference results from the fact that the scalar functions $\xi^A$ appearing
in expression (\ref{EQ5-1}) (which describe the profile of simple waves entering into a superposition) are substituted
in our case (see expressions (\ref{EQ3-3}) or (\ref{EQ3-4})) with a $q$ by $q$ or $k$ by $k$ matrix
$\Phi^1$ or $\Phi^2$, respectively.  This situation consequently allows for a much broader range of initial data.  The second
difference consists in fact that the restrictions (\ref{EQ5-5}) and (\ref{EQ5-8}) on the vector fields $\gamma_A$ and
$\lambda^A$, ensuring the solvability of the problem by GMC, are not necessary in our approach.  This makes it possible
for us to consider more general configurations of Riemann waves entering into an interaction.

A number of different attempts to generalize the Riemann
invariants method and its various applications can be found in the
recent literature on the subject (see e.g. \cite{Doy1,Fer2,Fer3,Fer4,Tsa1}).  For
instance, the nonlinear k-waves superposition
$u=f(r^1,\ldots,r^k)$ described in \cite{Pav1} can be regarded as
dispersionless analogues to "n-gap solution" of (\ref{EQ1-1})
which require the resolution of a set of commuting diagonal
systems for the Riemann invariants $r^A$, i.e.
\begin{equation}
\label{EQ5-10}
r^A_{x^i}=\mu^A_{i(j)}(r)r^A_{x^j},\quad A=1,\ldots,k,\quad i\neq j = 1,\ldots,p.
\end{equation}
That specific technique involves differential constraints on the
functions ${\mu^A_{ij}}$ of the form \cite{Tsa1}
\begin{equation}
\label{EQ5-11} \frac{\p_j
{\mu^A_{i(j)}}}{\mu^A_{i(j)}-\mu^A_{j(i)}} = \frac{\p_j
u^B_{i(j)}}{\mu^B_{i(j)}-\mu^B_{j(i)}},\quad i\neq j,\quad A\neq
B=1,\ldots,k,
\end{equation}
no summation.  As in the case of Riemann k-waves if the system (\ref{EQ5-11}) is satisfied for the functions
$\mu^A_{ij}$ then the general integral of the system (\ref{EQ5-10}) can be obtained by solving system (\ref{EQ5-9}) with respect to the
variables $r^1,\ldots,r^k$.

In contrast, our proposed approach does not require the use of differential equations (\ref{EQ5-10}) and
therefore does not impose constraints on the functions $\mu^A_{ij}$ when the 1-forms $\lambda^A$ are linearly
independent
and $k<p$.

However, if one removes these assumptions and $\lambda^A$ can be
linearly dependent and $k \geq p$ then the approach presented in
\cite{Fer3} is a valuable one and provides an effective tool for
classification criterion of integrable systems.

In order to verify the efficiency of our approach we have used it for constructing rank-2 solutions of several examples
of hydrodynamic type systems.  The proposed approach proved to be a useful tool in the case of multi-dimensional
hydrodynamic type systems (\ref{EQ1-1}), since it leads to new interesting solutions.

The examples illustrating our method clearly demonstrate its usefulness as it has produced several new and interesting results.  Let us note that the outlined approach to rank-k solutions lends itself to numerous potential applications
which arise in physics, chemistry and biology.  It has to be stressed that for many physical systems, (e.g. nonlinear
field equations, Einstein's equations of general relativity and the equations of continuous media, etc) there have been
very few, if any, known examples of rank-k solutions.  The approach proposed here offers a new and promising way to
investigate and construct such type of solutions.\\

\noindent {\bf Acknowledgements}\\
\noindent The research reported in this paper was partially supported by research grants
from NSERC of Canada and also by FQRNT from Gouvernement du Qu{\'e}bec.

%\begin{footnotesize}
\bibliographystyle{unsrt}

\begin{thebibliography}{99}

%\bibitem{Boi1}  Boilat, G., La propagation des ondes, Gauthier-Villars, 1965

%\bibitem{Bur1}  Burnat, M. :  The method of Riemann invariants and its applications to the theory of plasticity. I,II.Arch. Mech. (Arch. Mech. Stos.) {\bf 23}, 817-838 (1971); ibid {\bf 24}, 3-26 (1972)

%\bibitem{Bur2}  Burnat, M.:  The method of Riemann invariants for multi-dimensional nonelliptic system.  Bull. Acad. Polon. Sci. Sr. Sci. Tech. {\bf 17}, 1019-1026 (1969)

\bibitem{Bur3}  Burnat, M. : The method of characteristics and Riemann invariants for multidimensional hyperbolic systems. (Russian) Sibirsk. Mat. Z. {\bf 11}, 279-309 (1970)

%\bibitem{Cop1}  Copson E.T., "Partial differential equations," Cambridge Univ. Press, London, 1975

\bibitem{Cou1}  Courant R., Hilbert D., "Methods of mathematical physics",  Vol 1 and 2, Interscience, New York, 1962.

%\bibitem{Cou2}  Courant R., Friedrichs K.O., "Supersonic flow and shock waves," Intercience Publ., New York, 1948

%\bibitem{Cou3}  Courant R., Lax P., {\it On nonlinear partial differential equations with two independent variables,} Comm. Pure Appl. Math {\bf 2}, 255-273 (1949)

\bibitem{Daf1}  Dafermos, C. : {\it Hyperbolic conservation laws in continuum physics}.  Berlin-Heidelberg-New York:
Springer-Verlag, 2000

%\bibitem{Din1}  Dinu, L. : Some remarks concerning the Riemann invariance, Burnat-Peradzy{\'n}ski and Martin approaches. Rev. Roumaine Math. Pures Appl. {\bf 35}, 203-234 (1990)

%\bibitem{Dou1}  Douglis A., {\it Some existence theorems for hyperbolic systems of partial differential equations in two independent variables}, Comm. Pure Appl. Math. {\bf 5}, 119-154 (1952)

\bibitem{Doy1}  Doyle,P.W. and  Grundland,A.M.,  Simple waves and invariant solutions of quasilinear systems, J. Math. Phys. 37,6, 2969-2979 (1996)

\bibitem{Dub1}  Dubrovin, B.A.: Geometry of 2D topological field theories.  Lect. Notes in Math. {\bf 1620}, Berlin-Heidelberg New York: Springer-Verlag, 1996, pp.120-348

%\bibitem{Fer1}  Ferapontov, E.V., and K.R. Khusnutdinova,  On the integrability of (2+1)-Dimensional Quasilinear Systems, Com. Math. Phys 248, (2004), 187-206

\bibitem{Fer2}  Ferapontov, E.V., Pavlov, M.V. : Hydrodynamic reductions of the heavenly equation Class. Quantum Grav. {\bf 20}, 1-13, (2003)

\bibitem{Fer3}  Ferapontov, E.V., Khusnutdinova, K.R., On the integrability of (2+1)-dimensional quasilinear systems, Com. Math. Phys. 248, 187-206 (2004)

\bibitem{Fer4}  Ferapontov,E.V. , Khusnutdinova, K.R., The characterization of two-component (2+1)-dimensional integrable systems of hydrodynamic type, J. Phys. A : Math Gen 37, 2949-2963 (2004)

%\bibitem{Fri1}  Friedrichs K.O., {\it Nonlinear hyperbolic differential equations for functions of two independent variables}, Amer. J. Math. {\bf 70}, 555-589, 1948

\bibitem{Joh1}  John F., {\it Formulation of singularities in one-dimensional nonlinear wave propagation}, Comm. Pure Appl. Math. {\bf 27}, 377-405 (1974)

%\bibitem{God1}  Godunov,S.K. , An interesting class of quasilinear systems Dokl. Akad. Nauk SSSR 139,521-523 (1961)

\bibitem{Gru1}  Grundland,A.M.,  and P. Vassiliou, On the solvability of the Cauchy problem for Riemann double-waves by the Monge-Darboux method, Int. J Analysis 11, (1991), 221-278

%\bibitem{Gru2}  Grundland A.M., {\it Riemann invariants for nonhomogeneous systems of quasilinear partial differential equations. Conditions of involution}, Bull. Acad. Polon. Sci., Ser Sci Techn., Vol {\bf 22}, 4 (1974)

\bibitem{Gru3}  Grundland A.M., Zelazny R.,  {\it Simple waves in quasilinear hyperbolic systems, Part I and II}, J. Math Phys., Vol {\bf 24}, 9 (1983)

%\bibitem{Gru5}  Grundland A.M., Zelazny R., {\it Simple waves and their interactions in quasilinear hyperbolic systems}, Polish Acad. Sc.,Ed R. Teisseyre,  Publ. Inst. phys A-14, 1982, pp. 1-109

\bibitem{Gru4}  Grundland A.M. and Tafel J.,  Nonclassical symmetry reduction and Riemann wave solutions, J. Math Anal. Appl. 198, (1996) 879-892

%\bibitem{Gru6}  Grundland,A.M., J. Tafel, On the existence of nonclassical symmetries of partial diffeerntial equations, J. Math. Phys., 36, 3, 1425-1434 (1995)

%\bibitem{Har1}  Hartman P., Winter A., {\it On hyperbolic partial differential equations}, Amer. J. Math. {\bf 54}, 834-864, (1952)

%\bibitem{Jef1}  Jeffrey A., "Quasilinear hyperbolic systems and wave propagation," Pitman Publ., 1976

%\bibitem{Lig1}  Lighthill H., "Hyperbolic equations and waves," Springer-Verlag, New York, 1968

\bibitem{Maj1}  Majda A.,  "Compressible fluid flow and systems of conservation laws in several space variables, " Springer-Verlag, New York 1984

\bibitem{Olv2}  Olver, P.J., Applications of Lie groups to differential equations, Graduate Text in Math. 107, Springer-Verlag, New-York, 1986

\bibitem{Olv1}  Olver, P.J., Vorobev E.M., Nonclassical and conditional symmetries, CRC Handbook of Lie Groups Analysis, Vol 3, chapt. XI, CRC Press, London, Ed. N.H. Ibragimov, 1995

\bibitem{Pav1}  Pavlov,M.V,  Integrable hydrodynamic chains, J. Math. Phys. 44, 4139-4149 (2003)

\bibitem{Per1}  Peradzynski,Z. , Geometry of interactions of Riemann waves, in Advances in Nonlinear waves, Vol 3, Ed.Lokenath Debnath, Research Notes in Math 111, Pitman Advances Publ., Boston, 1985

\bibitem{Per2}  Peradzynski, Z., {\it Nonlinear plane k-waves and Riemann invariants}, Bull. Acad. Polon. Sci., Ser Sci Techn., Vol {\bf 19}, 9 (1971)

%\bibitem{Per3}  Peradzynski Z., {\it Riemann invariants for the nonplanar k-waves}, Bull. Acad. Polon. Sci., Ser Sci Techn., Vol {\bf 19}, 10 (1971)

\bibitem{Poi1}  Poisson S.D., {\it M{\'e}moire sur la th{\'e}orie du son}, in "Journal de l'{\'e}cole polytechnique, 14i{\`e}me cahier, 7, Paris, pp. 319-392, (1808)

\bibitem{Rie1}  Riemann B., {\it Uber die Fortpflanzung ebener Luftwellen von endlicher Schwingungsweite}, G{\"o}ttingen Abhandlungen, voll viii p.43 (Werke, 2te Aufl., Leipzig, P.157, 1892) (1858)

\bibitem{Roz1}  Rozdestvenskii B., Janenko N., {\it Systems of quasilinear equations and their applications to gas dynamics}, A.M.S., Vol {\bf 55}, Providence, 1983

%\bibitem{Roz2}  Rozdestvenskii B, Sidorenko A., {\it On the impossibility of gradient catastrophe for weakly nonlinear systems}, Comp. Math. and Math. Phys. {\bf 7} (1967)

%\bibitem{Ser1}  Serre,D. , Systems of conservation laws : 1. Hyperbolicity, entropies, shock waves, Cambridge University Press, 1999

\bibitem{Tsa1}  Tsarev, S.P.:  Geometry of hamiltonian systems of hydrodynamic type.  Generalized hodograph method. Izvestija AN USSR Math. {\bf 54}(5), 1048-1068 (1990)

%\bibitem{Whi1}  Whitham G.B., "Linear and nonlinear waves," John-Willey Pub., New York, 1974

%\bibitem{Zaj2}  Zajaczkowski W., {\it Riemann invariants interaction for nonelliptic systems},  Demonstratio Mathematica, {\bf 13}, 1,7-21 (1980)























\end{thebibliography}

%\end{footnotesize}

\clearpage

\clearpage

\end{document}